\documentclass[final,1p,times]{elsarticle}

\newcommand{\be}{\begin{equation}}
\newcommand{\ee}{\end{equation}}
\newcommand{\ba}{\begin{eqnarray}}
\newcommand{\ea}{\end{eqnarray}}

\usepackage{graphicx,amssymb,amsmath,color}

\journal{Nuclear Physics A}
\biboptions{sort&compress}
\begin{document}

\begin{frontmatter}

\title{Nucleon momentum distribution extracted from the experimental scaling function}

\author[label1]{M.~V.~Ivanov\corref{cor}}
\cortext[cor]{Corresponding author.}
\ead{martin.inrne@gmail.com}
\author[label1]{A.~N.~Antonov}
\author[label2]{J.~A. Caballero}

\address[label1]{Institute for Nuclear Research and Nuclear Energy, Bulgarian Academy of Sciences, Sofia 1784, Bulgaria}
\address[label2]{Departamento de F\'{i}sica At\'omica, Molecular y Nuclear, Universidad de Sevilla, 41080 Sevilla, Spain}

\begin{abstract}

The connection between the scaling function, directly extracted from the analysis of electron scattering data, and the nuclear spectral function or nuclear momentum density is investigated at depth. The dependence of the scaling function on the two independent variables in the scattering process, the transfer momentum ($q$) and the scaling variable ($y$), is taken into account, and the analysis is extended to both, positive and negative $y$-values, \textit{i.e.}, below and above the center of the quasielastic peak, respectively. Analytical expressions for the derivatives of the scaling function, evaluated at the finite limits of integration dealing with the kinematically allowed region, are connected with the spectral function. Here, contributions corresponding to zero and finite excitation energies are included. The scaling function is described by the Gumbel density distribution, whereas short-range correlations are incorporated in the spectral function by using some simple models. Also different parametrizations for the nucleon momentum distribution, that are compatible with the general properties of the scaling function, have been considered.

\end{abstract}


\end{frontmatter}

\section{Introduction\label{sec:introduction}}

One of the most fruitful concepts in Nuclear Physics is the nuclear spectral function $S(p_m,E_m)$. This gives the joint probability of finding a nucleon in a nucleus with given momentum (known as missing momentum $p_m$) and with a given excitation energy of the residual nuclear system (called the missing energy $E_m$). In general the spectral function is a complicated function whose evaluation requires a good knowledge of the many-body $A$-target and $(A-1)$-residual nuclear systems~\cite{PhysRevC.77.044311, Benhar:2015ula, Benhar:1994hw, Rocco16, Sick:1994vj, PhysRevC.97.035506}. The nuclear momentum density distribution $n(p_m)$ is obtained from the integral of the spectral function over the whole range of missing energy values. In general this would require to have available precise microscopic models of the interacting nuclear system for large values of the missing energy. Realistic nuclear models, that take into account the effects of nucleon-nucleon (NN) correlations, lead to non-vanishing contributions to the spectral function corresponding to more complex states with at least one of the spectator nucleons excited to the continuum~\cite{PhysRevC.83.045504, PhysRevC.53.1689, PhysRevC.41.R2474, PhysRevC.77.044311, Antonov:1994ty}. Calculations carried out for a variety of nuclear systems suggest that these contributions, arising from short-range dynamics, are nearly independent of the mass number, \textit{i.e.}, similar for all nuclei. It is important to point out that the correlated strength at high missing momentum, due to short-range-correlations (SRC), only appears at high missing energy~\cite{PhysRevC.83.045504, Donnelly:2017aaa, Ivanov:2013saa}.

We would like to note that the Green function method (GFM) [\emph{e.g.}~\cite{GROSS1970449, Frullani:1984nn, JEUKENNE197683, PhysRevC.99.025502}, see also~\cite{Antonov:1994ty} and~\cite{Giusti_2020}] is appropriate for the consideration of the spectral function in nuclear theory. Many difficulties arising in the case of finite nuclei can be avoided using the GFM in the case of nuclear matter. The results are applied to finite systems by introducing appropriate variables. Due to the fact that the properties of the hole distributions do not depend essentially on the details of nuclear structure, this procedure turns out to be quite convenient~\cite{Frullani:1984nn}. For instance, in~\cite{PhysRevC.97.035506} hole- and particle- spectral functions obtained within two distinct many-body methods are widely used to describe electroweak reactions in nuclei. As examples, we will note also the calculations of the GFM spectral function in Ref.~\cite{ORLAND1978442}, as well as the studies of influence of short-range correlations on the spectral function within the GFM in~\cite{ramos_spectral_1989} by summing the ladder diagrams in the perturbative expansion of the effective interaction.

The previous discussion clearly shows the difficulty in getting a precise description of the nuclear spectral function valid in a wide range of $(p_m,E_m)$-values. In most of the cases, one should rely on the information provided by different types of experiments, even if they are not capable of providing the spectral function for missing energies above some finite value~\cite{Donnelly:2017aaa}. This is the case of inclusive and semi-inclusive electron scattering reactions from nuclei, that have been largely used to constrain the spectral function (momentum distribution) in the missing-energy region corresponding to the contribution of the single-particle shell structure. In particular, the analysis of $(e,e'p)$ experiments has proved the validity of the shell structure of the nucleus, providing very precise information on the reduced cross section (identified with the momentum distribution) for the specific single-particle states~\cite{book01, Boffi:1993gs, Kelly2002, Frullani:1984nn, Udias:1996iy, Udias:1993zs, Udias:1993xy, Udias:1999tm}. These results have led some authors~\cite{Vagnoni:2017hll, Benhar:2015wva} to construct realistic spectral functions by using information on the $(e,e'p)$ data at low-missing energies and different models of the interacting nuclear system for larger $E_m$-values. Thus the general expression for the spectral function is divided in two terms, $S(p_m,E_m)=S_{\mathrm{IPSM}}(p_m,E_m)+S_{\mathrm{corr}}(p_m,E_m)$, the former given by an independent particle shell model (IPSM) approximation but with the individual shells widened using different functions, like Lorentzians, whereas the latter, $S_{\mathrm{corr}}$, connected to contributions ascribed to NN correlations.

In addition to semi-inclusive, $(e,e'p)$, processes, also inclusive electron scattering can provide useful information on the total momentum distribution of nuclei. The analysis of quasielastic (QE) $(e,e')$ reactions leads to the phenomenon of scaling, \textit{i.e.}, the scaling function defined as the differential $(e,e')$ cross section divided by an appropriate factor including the single-nucleon cross section, is shown to depend only on a single variable ($y$-scaling variable), given as a particular combination of the energy ($\omega$) and momentum ($q$) transferred to the nucleus~\cite{PhysRevC.38.1801, PhysRevLett.82.3212, PhysRevC.60.065502, Maieron:2001it, Caballero05, Caballero:2006wi}. This is known as scaling of first kind, that is, independence on $q$. Scaling of second kind refers to the scaling function being independent on the mass number. The existence of both types of scaling, that occurs at excitation energies below the QE peak, is denoted as superscaling. As shown in \cite{Maieron:2001it}, the analysis of the isolated longitudinal $(e,e')$ data leads to an universal scaling function that, when plotted against the superscaling variable (denoted as $\psi$), presents a significative asymmetry with a long tail extended to large-positive values of $\psi$ (region above the QE peak). The reader interested in a detailed discussion of the phenomenon of scaling, also extended to the region of high nucleon resonances, can go to Refs.~\cite{Amaro:2006if, Caballero:2007tz, Amaro:2005dn, Barbaro:2003ie, Maieron09}.

Here we would like to add also the analysis in~\cite{PhysRevC.96.015504} of the scaling properties of the electromagnetic response function of $^4$He and $^{12}$C nuclei within the Green's function Monte-Carlo (GFMC) approach~\cite{RevModPhys.87.1067} using only one-body current contributions. The mentioned two approaches in~\cite{PhysRevC.97.035506} lead to compatible nucleon-density scaling functions that for large momentum transfers satisfy first-kind of scaling and has an asymetric shape. The formalism used in~\cite{PhysRevC.97.035506} based on the impulse approximation combines a fully relativistic description of the electromagnetic interaction with a treatment of nuclear dynamics in the initial state. As noted in~\cite{PhysRevC.97.035506} the final state interactions (FSI) are treated as corrections that requires further approximation~\cite{PhysRevC.87.024606, PhysRevD.91.033005}.

In the Relativistic Green's function (RGF) model FSI were originally developed within a nonrelativistic~\cite{Capuzzi:1991qd, Capuzzi:2004au} and then within a fully relativistic framework~\cite{Meucci:2003uy, Meucci:2005pk} for the inclusive quasielastic (QE) electron scattering using complex energy-dependent optical potential. The model was successfully applied to electron scattering data~\cite{Capuzzi:1991qd, Capuzzi:2004au, Meucci:2003uy, Boffi:1993gs, Meucci:2009nm, esotici2} and later extended to neutrino-nucleus scattering~\cite{Meucci:2003cv, Meucci:2014pka, Ivanov:2016pon}.

In this work our main interest is centered in the connection between the scaling (superscaling) function, extracted directly from the analysis of scattering data, and the spectral function (momentum distribution), evaluated using different nuclear models. As already presented in some previous works~\cite{PhysRevC.81.055502,PhysRevC.83.045504}, this connection only emerges in a clear way under certain restrictive approximations considered in the description of the scattering reaction formalism. In particular, all studies of electron scattering reactions making use of the spectral function are based on the {\it factorization ansatz}, \textit{i.e.}, the $(e,e')$ cross section factorizes into two terms: one dealing with the single-nucleon-photon vertex (single-nucleon responses) and the other containing the whole information about the nuclear systems involved in the process (nuclear spectral function). This factorization result breaks in general. Even in the impulse approximation, namely, only one single-nucleon being active in the scattering process (the other nucleons treated as mere spectators), the effects introduced by the final state interactions (FSI) between the ejected nucleon and the residual nucleus, and/or the role played by the lower components in the relativistic nucleon wave functions, break factorization at some level. Notwithstanding, the scaling (superscaling) behavior shown by data proves unambiguously that the differential $(e,e')$ cross section being given as ``single-nucleon'' cross sections times a universal scaling function results an excellent approximation in certain kinematical domains, namely, high values of the transfer momentum where the QE reaction mechanism is dominant.

The connection between the scaling (superscaling) function and the nuclear spectral function (or momentum distribution) still deserves some discussion. Notice that the argument can be applied in a twofold way. First one can develop theoretical nuclear models, with a high level of complexity, providing a realistic nuclear spectral function to be used in the analysis and interpretation of electron scattering data, \textit{i.e.}, the scaling/superscaling function.  A different strategy is to use directly as input the data extracted from the experiment, namely the scaling function, in order to get precise information on the spectral function, and in particular, on the effects associated to NN correlations. In this work we follow this second procedure. We do not intend to provide ``sophisticated'' theoretical descriptions of the spectral function, but instead, use our present knowledge on the scaling function and its explicit dependence with the missing energy and momentum in order to find out a precise connection between the derivatives of the scaling function and the behavior of the momentum density. We already followed this general strategy in the past, but making use of some restricted approximations~\cite{PhysRevC.81.055502} or specific simple nuclear models based on the independent particle approach~\cite{PhysRevC.83.045504}. Some other authors~\cite{PhysRevC.53.1689, CiofidegliAtti:182876, ATTI1989361, PhysRevC.43.1155, PhysRevC.39.259, DEGLIATTI1987127, PhysRevC.41.R2474} have also used a similar procedure by isolating in the spectral function the term that yields the probability distribution that the final $(A-1)$ system is left in any of its excited states. This ``excited'' spectral function is later connected with ``binding corrections'' to the scaling function. In this paper we follow a similar strategy, but solving exactly the various integro-differential equations that connect the spectral function (momentum distribution) with the scaling (superscaling) functions. In doing so, we extend our previous work in~\cite{PhysRevC.81.055502} by explicitly accounting for effects coming from the NN correlations.

The general structure of the paper is as follows. In Section~\ref{sec:formalism} we revisit the general formalism by discussing in detail the connection between the scaling and the spectral function. We summarize the most relevant results connected with scaling arguments and introduce the functions and variables of interest for the general discussion. We also present the basic equations related to the analysis of the spectral function and/or the momentum distribution, and discuss the results obtained. In Section~\ref{sec:results} we focus on the momentum density derived from the scaling function. Here we discuss in detail the results obtained with particular emphasis in the Gumbel distribution. Finally, Section~\ref{sec:conclusions} summarizes the main conclusions of this work.

\section{General Formalism\label{sec:formalism}}

\subsection{Scaling Function \emph{vs.} Nucleon Momentum Distribution\label{subsec21}}

Within the Plane-Wave Impulse Approximation (PWIA) the ($e,e'N$) differential cross section factorizes in two basic
terms~\cite{Frullani:1984nn, Boffi:1993gs, book01, Kelly2002}:
\be
\left[\frac{d\sigma}{d\epsilon'd\Omega'dp_Nd\Omega_N} \right]_{(e,e'N)}^{\text{PWIA}}= K\sigma^{eN}(q,\omega;p,{\cal E},\phi_N)S(p,{\cal E})\,,\label{eq:PWIA}
\ee
the electron-nucleon cross section for a moving, off-shell nucleon, $\sigma^{eN}$, and the spectral function $S(p,{\cal E})$ that gives the combined probability that, after a nucleon with momentum $p$ has been removed from the target, the $(A-1)$-nucleon system is left with excitation energy ${\cal E}$. In Eq.~(\ref{eq:PWIA}) $K$ is a kinematical factor, $p\equiv p_m$ is the missing momentum and ${\cal E}$ is the excitation energy that is essentially the missing energy minus the separation energy. Further
assumptions are necessary~\cite{PhysRevC.81.055502} to show how the scaling function $F(q,\omega )$ emerges from the PWIA, namely, the spectral function is assumed to be isospin independent and $\sigma ^{eN}$ is assumed to have a very mild dependence on $p$ and $\cal E$. Hence the $eN$ cross section can be evaluated at fixed values of $p$ and ${\cal E}$ and typically the differential cross section for inclusive QE $(e,e')$ processes is written in the form~\cite{PhysRevC.81.055502, PhysRevC.83.045504}:
\be
\left[\frac{d\sigma}{d\epsilon'd\Omega'}\right]_{(e,e')}\cong \overline{\sigma}^{eN}(q,\omega;p=|y|,{\cal E}=0)\cdot F(q,\omega) \, , \label{eq:scaling}
\ee
where the single-nucleon cross section is evaluated at the special kinematics $p=|y|$ (with the scaling variable $y$ being the lowest longitudinal momentum of a bound nucleon when the residual nucleus is in its ground state, ${\cal E} = 0$~\cite{DEGLIATTI1987127}; see also the next Section). This corresponds to the lowest value of the missing momentum occurring when ${\cal E}=0$. The term $\overline{\sigma}^{eN}$ refers to the azimuthal-angle-averaged single-nucleon cross section and it also incorporates the contribution of all nucleons in the target.

The function $F(q,\omega)$ in Eq.~(\ref{eq:scaling}) is known as the scaling function and it is given in PWIA in terms of the spectral function:
\be
F(q,\omega)= 2\pi\int\!\!\!\int_{\Sigma(q,\omega)}p\,dp\, d{\cal E}\,S(p,{\cal E}) \, , \label{eq:scaling_function}
\ee
where ${\Sigma(q,\omega)}$ represents the kinematically allowed region, $p$ is the struck nucleon's momentum and
\be
{\cal E}(p)\equiv \sqrt{M_B^{*^2}+p^2}-\sqrt{M_B^{0^2}+p^2} \geq 0 \, ,
\ee
the excitation energy of the recoiling system $B$, with $M_B^0$ the ground-state mass of the residual nucleus and $M_B^*$ the general  invariant mass of the daughter final state. The integration in Eq.~(\ref{eq:scaling_function}) is extended to the kinematically allowed region in the $(p,{\cal E})$ plane at fixed values of the momentum and energy transfer, $(q,\omega)$. The general kinematics corresponding to QE $(e,e')$ processes leads to the following ${\cal E}$-integration range
\be
\max\{0,{\cal E}^+\}\leq{\cal E}\leq{\cal E}^- \, ,
\ee
where
\be
{\cal E}^{\pm}(p;q,\omega)=(M_A^0+\omega)-\left[\sqrt{(q\pm p)^2+m_N^2}+ \sqrt{M_B^{0^2}+p^2}\right]\label{eq:exc}
\ee
and where $M_A^0$ is the target nuclear mass and $m_N$ the nucleon mass.

\begin{figure}[htb]
\centering\includegraphics[width=0.8\textwidth]{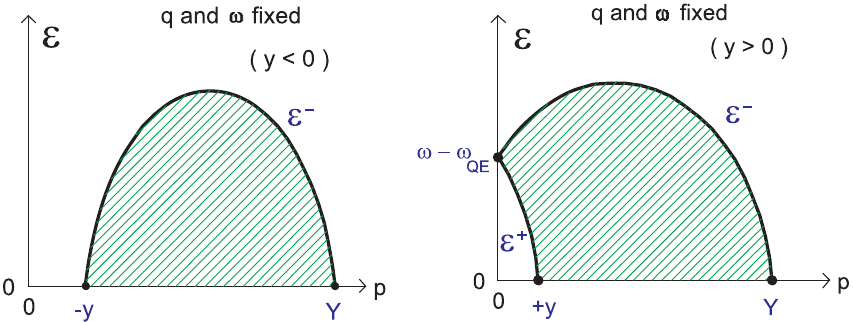}
\caption{(Color online) Excitation energy corresponding to negative (left) and positive (right) values of $y$.\label{fig:regions}}
\end{figure}

The integration region $\Sigma(q,\omega)$  is shown in Fig.~\ref{fig:regions} for fixed values of the transferred energy and momentum for $\omega<\omega_{QE}$ (left-hand panel) and $\omega>\omega_{QE}$ (right-hand panel), with $\omega_{QE}$ the energy where the quasielastic peak (QEP) occurs. The intercepts between the curve ${\cal E}^-$ and the $p$-axis will be denoted by $-y$ and $Y$, {\it i.e.,} ${\cal E}^-(-y;q,\omega)={\cal E}^-(Y;q,\omega)=0$. In the region below the QEP, $y$ is negative and $p=-y$ represents the minimum value for the struck nucleon's momentum. Above the QEP $y$ is positive and the curve ${\cal E}^+$ cuts the integration region when $p<y$.

Using as independent variables $(p,{\cal E};q,y)$, the energy transfer can be expressed as:
\be
\omega(q,y) = \sqrt{(q+y)^2+m_N^2}+\sqrt{M_B^{0^2}+y^2}-M_A^0 \, , \label{eq:omega}
\ee
the limits of the excitation energy
\be
{\cal E}^\pm (p;q,y) = \left[\sqrt{(q+y)^2+m_N^2}-\sqrt{(q\pm p)^2+m_N^2} \right]+\left[\sqrt{M_B^{0^2}+y^2}-\sqrt{M_B^{0^2}+p^2}\right] \label{eq:exc1}
\ee
and the upper limit of $p$:
\begin{equation}
Y(q,y) = \frac{ M_B^{0^2}(2q+y) + 2(q+y)\sqrt{M_B^{0^2}+y^2}\sqrt{(q+y)^2+m_N^2} +y\left[2(q+y)^2+m_N^2\right] }
{M_B^{0^2}+2\sqrt{M_B^{0^2}+y^2}\sqrt{(q+y)^2+m_N^2} + 2y(q+y)+m_N^2} \, . \label{eq:Ylarge_app}
\end{equation}

Then the scaling function in Eq.~(\ref{eq:scaling_function}) can be recast as follows
\begin{equation}
{\dfrac{1}{2\pi}}F(q,y) = \int\limits_{-y}^{Y(q,y)}p\,dp\int\limits_0^{{\cal E}^-(p;q,y)}d{\cal E}
S(p,{\cal E}),\ \ \mbox{if}\ y<0, \label{eq:Fneg}
\end{equation}
\begin{equation}
{\dfrac{1}{2\pi}}F(q,y) = \int\limits_{0}^y p\,dp\int\limits_{{\cal E}^+(p;q,y)}^{{\cal E}^-(p;q,y)}
d{\cal E} S(p,{\cal E})
+ \int\limits_{y}^{Y(q,y)}p\,dp\int\limits_0^{{\cal E}^-(p;q,y)}d{\cal E} S(p,{\cal E}),
\ \ \mbox{if}\ y>0, \label{eq:Fpos}
\end{equation}
for negative and positive values of $y$, respectively.

In what follows we will consider our analysis in relation with the scaling method developed by C. Ciofi degli Atti \emph{et al.} (see, \emph{e.g.}~\cite{PhysRevC.53.1689, PhysRevC.41.R2474, CiofidegliAtti:182876, CIOFIDEGLIATTI1990349, ATTI1989361, PhysRevC.43.1155, 10.1007/978-3-7091-8897-2_30, PhysRevC.36.1208, PhysRevC.39.259, DEGLIATTI1987127}). It was shown in the latter that information on the nucleon momentum distribution can be extracted from the inclusive ($e,e'$) cross sections and that within the PWIA the inelastic cross section depends on nuclear structure peculiarities through the spectral function $S(p,{\cal E})$ [the notations in the mentioned papers are usually $P(k,E)$ for the spectral function with components $P_{gr}(k,E)$ and $P_{ex}(k,E)$, where $P_{gr}(k,E)$ yields the probability distribution that the final $(A-1)$ system is left in its ground state and $P_{ex}(k,E)$ is that part of the spectral function which accounts for the excited states of the final $(A-1)$ system]. It turns out also that the inclusive cross section is not directly related to the nucleon momentum distribution but depends on the knowledge of the full spectral function $S(p,{\cal E})$. We note that the consideration in Refs.~\cite{CiofidegliAtti:182876, CIOFIDEGLIATTI1990349, ATTI1989361, PhysRevC.43.1155, 10.1007/978-3-7091-8897-2_30, PhysRevC.36.1208, PhysRevC.39.259, DEGLIATTI1987127} is restricted to the case of negative values of $y$. The nucleon momentum distribution of $^2$H was obtained from the $y$-scaling analysis of inclusive electron scattering in Ref.~\cite{PhysRevC.36.1208}. The method was extended for $^3$He (\emph{e.g.}~\cite{PhysRevC.39.259, DEGLIATTI1987127}), for $^3$He and $^4$He (\emph{e.g.}~\cite{CIOFIDEGLIATTI1990349}), and later it was applied to complex nuclei (\emph{e.g.}~\cite{PhysRevC.53.1689, PhysRevC.41.R2474, ATTI1989361, PhysRevC.43.1155}). It has been noted~\cite{PhysRevC.43.1155} that in the latter case ``it is also useful to adopt another representation of the spectral function in which the ground state of the $(A-1)$ system and its excited states represented by one-hole excitations are explicitly separated from more complex configurations, \emph{e.g.}, one-particle--two-hole states, which can be reached when two-particle--two-hole states in the target nucleus are considered''.

In Ref.~\cite{KULAGIN2006126} (see also Ref.~\cite{PhysRevC.74.054316}) was developed a method that leads to the approximate expression for the spectral function which incorporates both the single-particle nature of the spectrum at low- and high-energy and high-momentum components due to \emph{NN}-correlations in the ground state. The low-energy part is described by the mean-field spectral function for which the authors use an approximate expression motivated by closure (\emph{i.e.} the sum over occupied levels is substituted by its average value). This approach allowed us to apply in~\cite{PhysRevC.81.055502} the mentioned approximation that leads to splitting the spectral function into two terms, corresponding to zero and finite excitation energy, respectively:
\be \label{eq:spectral}
S(p,{\cal E}) = n_0(p)\delta({\cal E}) + S_1(p,{\cal E})
\ee
with $S_1(p,{\cal E}=0)=0$, which, inserted in Eqs.~(\ref{eq:Fneg}) and~(\ref{eq:Fpos}) yields
\be
{\dfrac{1}{2\pi}}F(q,y<0) = \int\limits_{-y}^{Y(q,y)}p\,dp\, n_0(p)+\int\limits_{-y}^{Y(q,y)} p\,dp\int\limits_0^{{\cal E}^-(p;q,y)}d{\cal E}S_1(p,{\cal E}),\label{eq:Fneg1}
\ee
\be
{\dfrac{1}{2\pi}}F(q,y>0) =
\int_{y}^{Y(q,y)}p\,dp\, n_0(p)+\left[ \int\limits_{0}^y p\,dp\int\limits_{{\cal E}^+(p;q,y)}^{{\cal E}^-(p;q,y)}d{\cal E} + \int\limits_{y}^{Y(q,y)}p\,dp\int\limits_0^{{\cal E}^-(p;q,y)}d{\cal E}
\right] S_1(p,{\cal E}) \, .\label{eq:Fpos1}
\ee

In order to analyze how the scaling function and the nucleon momentum distribution are connected, we proceed by evaluating the derivatives of the scaling function $F$ with respect to $y$ and $q$ making use of the Leibniz's formula and choosing $(p;q,y)$ as the three remaining independent variables. In this work, in contrast to Ref.~\cite{PhysRevC.81.055502}, we use the full expressions for the derivatives. In our calculations we assume that the scaling of the first kind is fulfilled (the scaling function $F$ loses its dependence upon $q$):
\be
\lim_{q\to\infty} \frac{\partial F}{\partial q} \simeq0 \quad \Rightarrow \quad
\lim_{q\to\infty} \frac{\partial F}{\partial q}\left(\frac{\partial Y}{\partial y}\right) \simeq0\,,\label{eq:sc.f.I_kind}
\ee
for negative and positive values of $y$. This is strictly valid only for very large values of $q$ (\emph{e.g.} at $q>500$~MeV) and it is entirely based on the approximations leading to the expression in Eq.~(\ref{eq:scaling_function}) that connects the scaling function to the spectral function. In~\ref{appendixC} we discuss in detail the validity of the approximations involved in Eq.~(\ref{eq:sc.f.I_kind}). After some algebra, presented in~\ref{appendixA}, we get the following results:

\subsubsection{Negative--$y$ region}

\be
n(k)={\dfrac{1}{2\pi k}} \left(\frac{\partial F}{\partial y}\right)_{y=-k} + \int\limits_{0}^{\infty} d{\cal E}\,S_1(k,\,{\cal E})\,-\,\dfrac{1}{k}\left[\int\limits_{-y}^{Y(q,y)} \mathcal{D}_1(p;q,y)\,S_1(p,\,{\cal E}^-)\,p\,dp\right]_{y=-k}, \label{eq:n_neg}
\ee
where
\be
\mathcal{D}_1(p;q,y)= \left( \frac{\partial{\cal E}^-}{\partial y} -\frac{\partial {\cal E}^-}{\partial q}
\frac{(\partial Y/\partial y)}{(\partial Y/\partial q)} \right) \, .\label{eq:D1}
\ee
Therefore, in the case of negative values of $y$ the momentum distribution $n_0(k)$ can be expressed as:
\be
n_0(k) = {\dfrac{1}{2\pi k}} \left(\frac{\partial F}{\partial y}\right)_{y=-k} -\, \dfrac{1}{k} \left[ \int\limits_{-y}^{Y(q,y)} \mathcal{D}_1(p;q,y)\, S_1(p,\,{\cal E}^-)\,p\,dp \right]_{y=-k}. \label{eq:n0_neg}
\ee

On this point we would like to note that in Refs.~\cite{CiofidegliAtti:182876, CIOFIDEGLIATTI1990349, ATTI1989361, PhysRevC.43.1155, 10.1007/978-3-7091-8897-2_30, PhysRevC.36.1208, PhysRevC.39.259, DEGLIATTI1987127} the scaling function has the following form in the asymptotic limit ($q \rightarrow \infty$):
\be
F(y)=f(y)-B(y)\,,\label{eq:ciofi01}
\ee
where
\be
f(y)=2\pi\int\limits_{|y|}^{\infty}n(k)\,k\,dk\label{eq:ciofi02}
\ee
and
\be
B(y)=2\pi\int\limits_{E_{\min}}^{\infty}dE\int\limits_{|y|}^{k_{\min}^{\infty}(y,E)}P_{ex}(k,E)\,k\,dk\,.\label{eq:ciofi03}
\ee
In Eq.~(\ref{eq:ciofi03}) the lower limit in the energy is given by $E_{\min}=|E_A|-|E_{A-1}|$, $E_A$ and $E_{A-1}$ being the ground state energies of the initial and final nuclei, and $P_{ex}(k,E)$ is the part of the spectral function corresponding to the spectator $A-1$ system in all possible virtual excited states. For large values of $q$
\be
k_{\min}^{\infty}(y,E)\cong|y-(E-E_{\min})|\,.\label{eq:ciofi05}
\ee
The quantity $B(y)$ causes the ``scaling violation'' due to the nucleon binding. Taking derivative of both sides of Eq.~(\ref{eq:ciofi01}) one gets:
\be
n(k)=-\dfrac{1}{2\pi y}\Bigg[\dfrac{dF(y)}{dy}+\dfrac{dB(y)}{dy}\Bigg]\,,\quad k=|y| \,.\label{eq:ciofi06}
\ee
It is noted in the works mentioned above that: i) the extraction of $n(k)$ in the approach (at $y<0$) needs the asymptotic scaling function $F(y)$ to be obtained from the experimental data, and ii) the binding correction term $dB/dy$ to be estimated in a realistic way.  We note that our method is a natural extension and development along this line. As can be seen, the comparison of Eq.~(\ref{eq:n_neg}) with Eq.~(\ref{eq:ciofi06}) gives the following correspondence of $dB/dy$ obtained in the approach followed by Ciofi degli Atti \emph{et al.} to the term obtained in our method:
\be
-\dfrac{1}{2\pi y}\dfrac{dB(y)}{dy}=\int\limits_0^{\infty}S_1(-y,{\cal E})\,d{\cal E}+ \dfrac{1}{y}
\int\limits_{-y}^{Y(q,y)} \mathcal{D}_1(p;q,y)\, S_1(p,\,{\cal E}^-)\,p\,dp \, . \label{eq:ciofi07}
\ee

We note that the right-hand side of Eq.~(\ref{eq:ciofi07}) gives additional and more complex information on the quantity $B(y)$, its relation to the correlated part of the momentum distribution and the kinematical conditions (energy, transferred momentum, the scaling momentum and others).

\subsubsection{Positive--$y$ region}

\begin{multline}
n(k) = - {\dfrac{1}{2\pi k}}\left(\frac{\partial F}{\partial y}\right)_{y=k}+\int\limits_{0}^{\infty} d{\cal E}\,S_1(k,\,{\cal E})
+\dfrac{1}{k}\left[ \int\limits_0^{Y(q,y)} \mathcal{D}_1(p;q,y)\,S_1(p,\,{\cal E}^-)\,p\,dp\right]_{y=k}  \\
-\dfrac{1}{k}\left[\int\limits_0^y  \mathcal{D}_2(p;q,y)\,S_1(p,\,{\cal E}^+)\,p\,dp\right]_{y=k}, \label{eq:n_pos}
\end{multline}
where
\be
\mathcal{D}_2(p;q,y)= \left( \frac{\partial{\cal E}^+}{\partial y} -\frac{\partial {\cal E}^+}{\partial q}
\frac{(\partial Y/\partial y)}{(\partial Y/\partial q)} \right) \, . \label{eq:D2}
\ee
Therefore, in the case of positive values of $y$ the momentum distribution $n_0(k)$ can be expressed as:
\begin{multline}
n_0(k) = - {\dfrac{1}{2\pi k}}\left(\frac{\partial F}{\partial y}\right)_{y=k} +\dfrac{1}{k}\left[
\int\limits_0^{Y(q,y)} \mathcal{D}_1(p;q,y)\,S_1(p,\,{\cal E}^-)\,p\,dp\right]_{y=k} \\
- \dfrac{1}{k}\left[\int\limits_0^y  \mathcal{D}_2(p;q,y)\,S_1(p,\,{\cal E}^+)\,p\,dp\right]_{y=k}.\label{eq:n0_pos}
\end{multline}

Comparing Eq.~(\ref{eq:n0_neg}) and Eq.~(\ref{eq:n0_pos}) for negative and positive values of $y$, we can write the following connection between the scaling function $F(y)$ and $S_1(p,{\cal E})$:
\begin{multline}
{\dfrac{1}{2\pi }} \left[ \left(\frac{\partial F}{\partial y}\right)_{y=-k}+ \left(\frac{\partial F}{\partial y}\right)_{y=k} \right] = \left[ \int\limits_{-y}^{Y(q,y)} \mathcal{D}_1(p;q,y)\, S_1(p,\,{\cal E}^-)\,p\,dp \right]_{y=-k}\\
+ \left[ \int\limits_0^{Y(q,y)} \mathcal{D}_1(p;q,y)\,S_1(p,\,{\cal E}^-)\,p\,dp\right]_{y=k} - \left[\int\limits_0^y  \mathcal{D}_2(p;q,y)\,S_1(p,\,{\cal E}^+)\,p\,dp\right]_{y=k}. \label{eq:ScF.vs.SF}
\end{multline}
In Section~\ref{sec:results} we present more details on the corresponding connection between the results in Eq.~(\ref{eq:ScF.vs.SF}) and the superscaling function. Here, we point out that if $S_1(p,{\cal E})=0$, as in the case of RFG, then it follows from~(\ref{eq:ScF.vs.SF}) that the scaling function is symmetric [$F(y)=F(-y)$].

\begin{figure}[tbh]
\centering\includegraphics[width=0.7\textwidth]{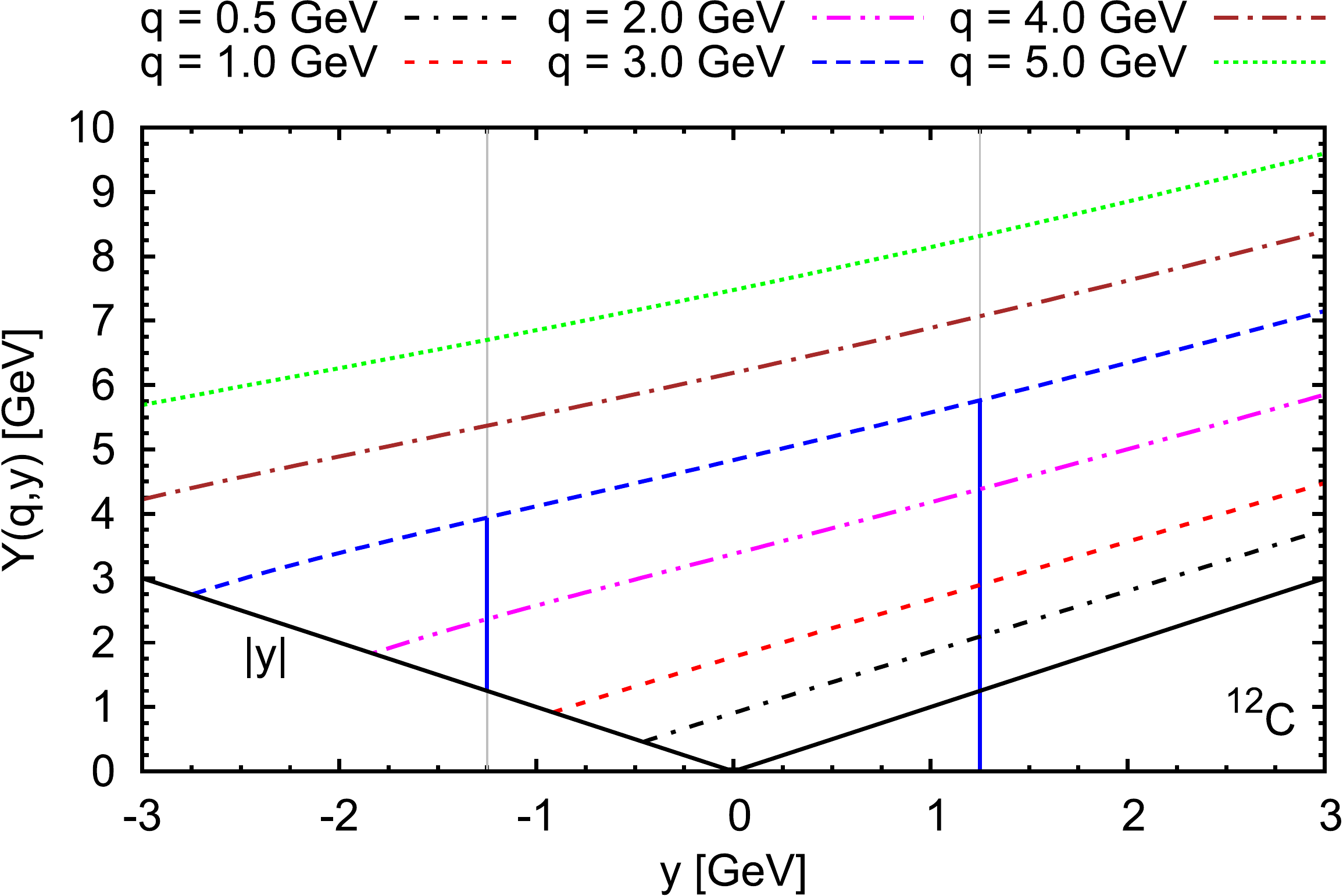}
\caption{(Color online) Results for the $Y(q,y)$ for $^{12}$C nucleus using analytical expression from Eq.~(\ref{eq:Ylarge_app}).\label{fig:Yc}}
\end{figure}

In the next several figures we present the main part of the variables which are used in Eqs.~(\ref{eq:n_neg})--(\ref{eq:ScF.vs.SF}) and their kinematical behaviour. In Fig.~\ref{fig:Yc} is shown $Y(q,y)$ (the upper limit of $p$) as a function of $y$ for several fixed values of $q$. Notice that $|y|$ is also shown by the black solid line. To make clearer the results, let us consider as an example the case shown by the blue dashed line, that is, $Y(q,y)$ as a function of $y$ at fixed value of $q=3$~GeV; the available values of $p$ at $y = -1.25$~GeV ($1.25$~GeV$\leq p \leq 3.94$~GeV) and $1.25$~GeV ($0$~GeV$\leq p \leq 5.76$~GeV) are presented by the vertical blue solid lines; $Y(q,y)= |y|$ at $y_0 \sim -2.7482$~GeV, for values of $y$ less than $y_0$: $Y(q,y)< |y|$ therefore this region is not allowed.

\begin{figure*}[htb]
\centering
\includegraphics[width=.495\textwidth]{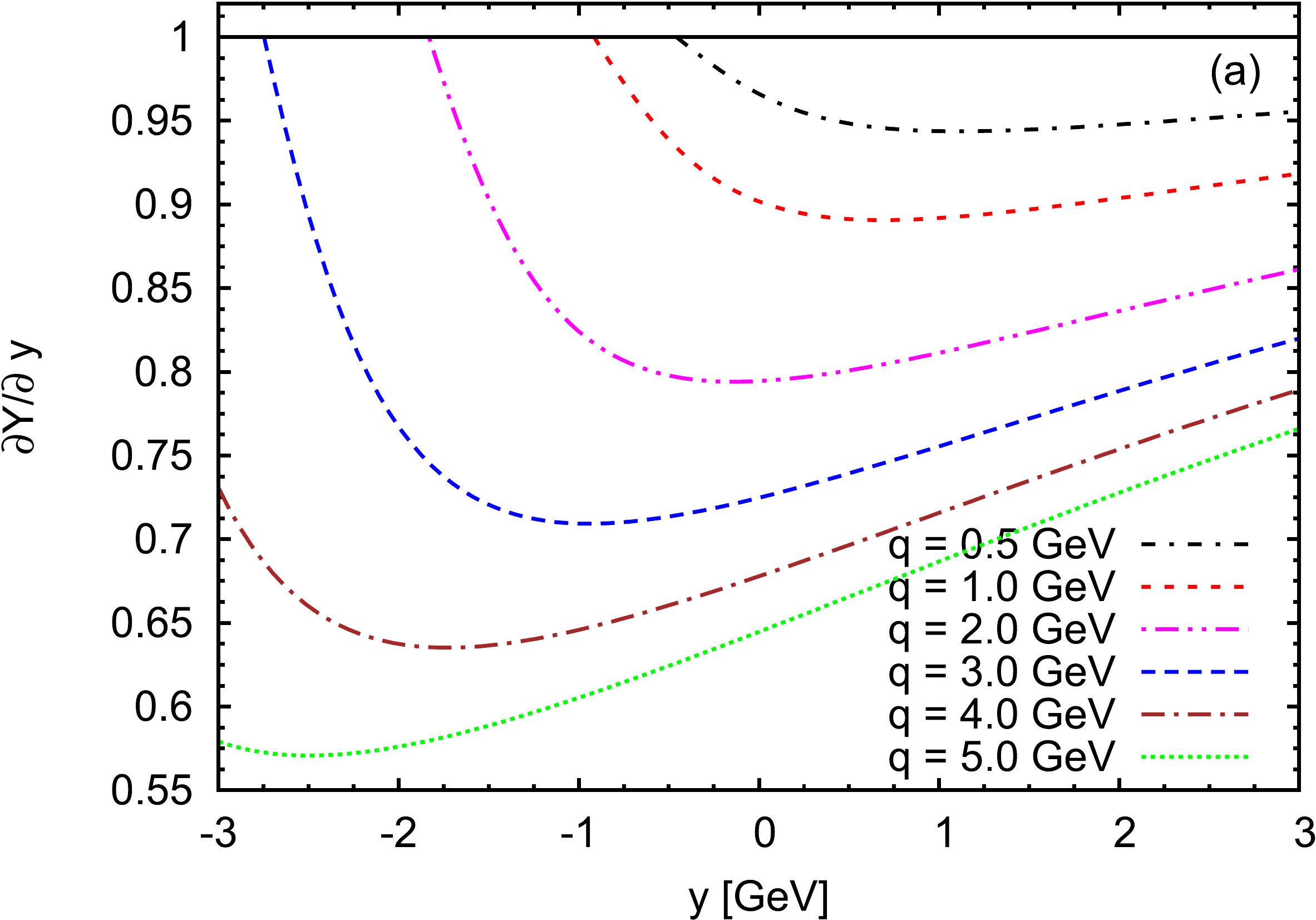}\hfill\includegraphics[width=.495\textwidth]{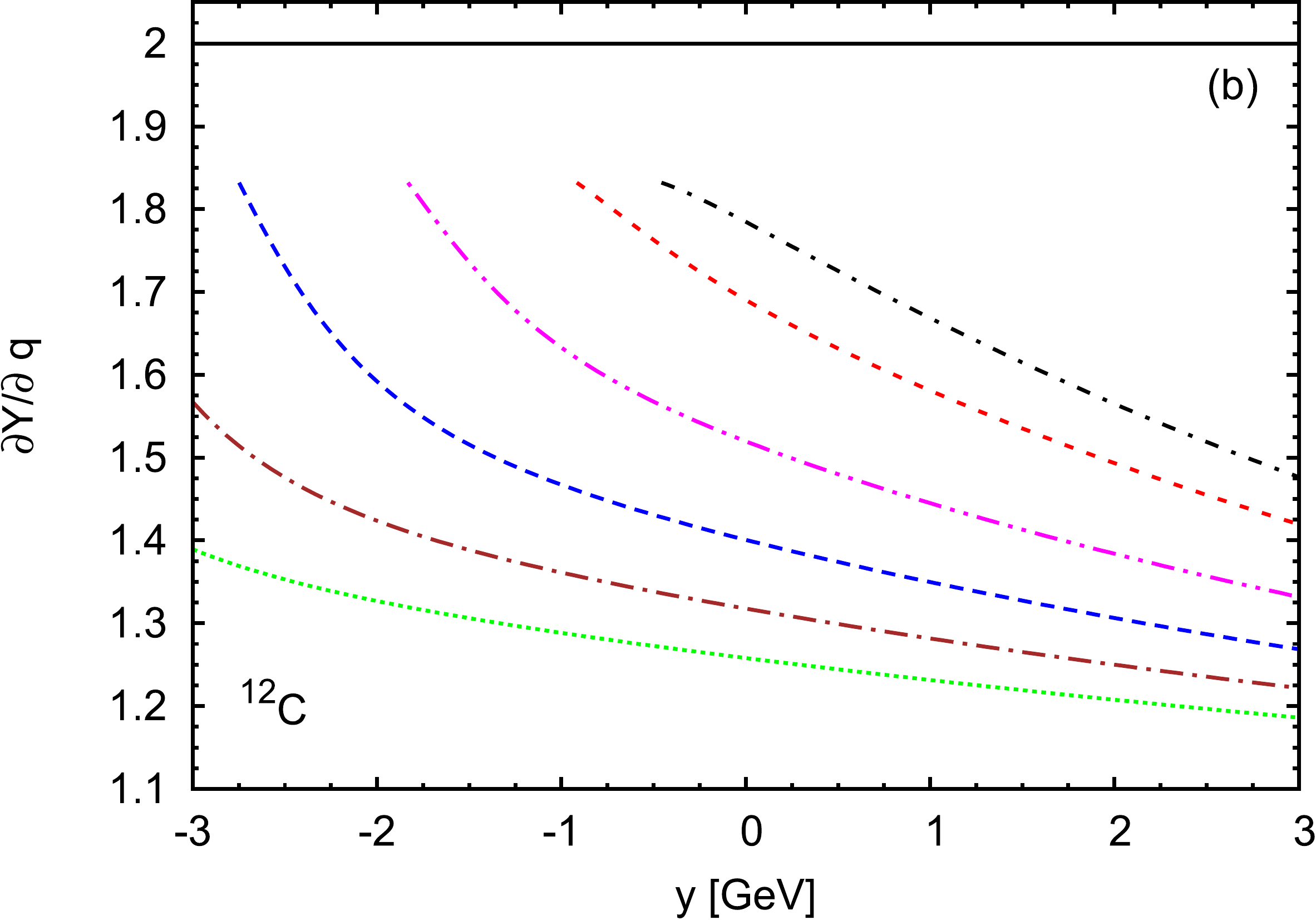}
\caption{(Color online) Results for the $\partial Y/\partial y$ [panel (a)] and $\partial Y/\partial q$ [panel (b)] for $^{12}$C nucleus using analytical expressions of derivatives obtained from Eq.~(\ref{eq:Ylarge_app}).
\label{fig:diffY_C12}}
\end{figure*}

\begin{figure*}[htb]
\begin{minipage}[t]{0.49\textwidth}
\centering\includegraphics[width=\textwidth]{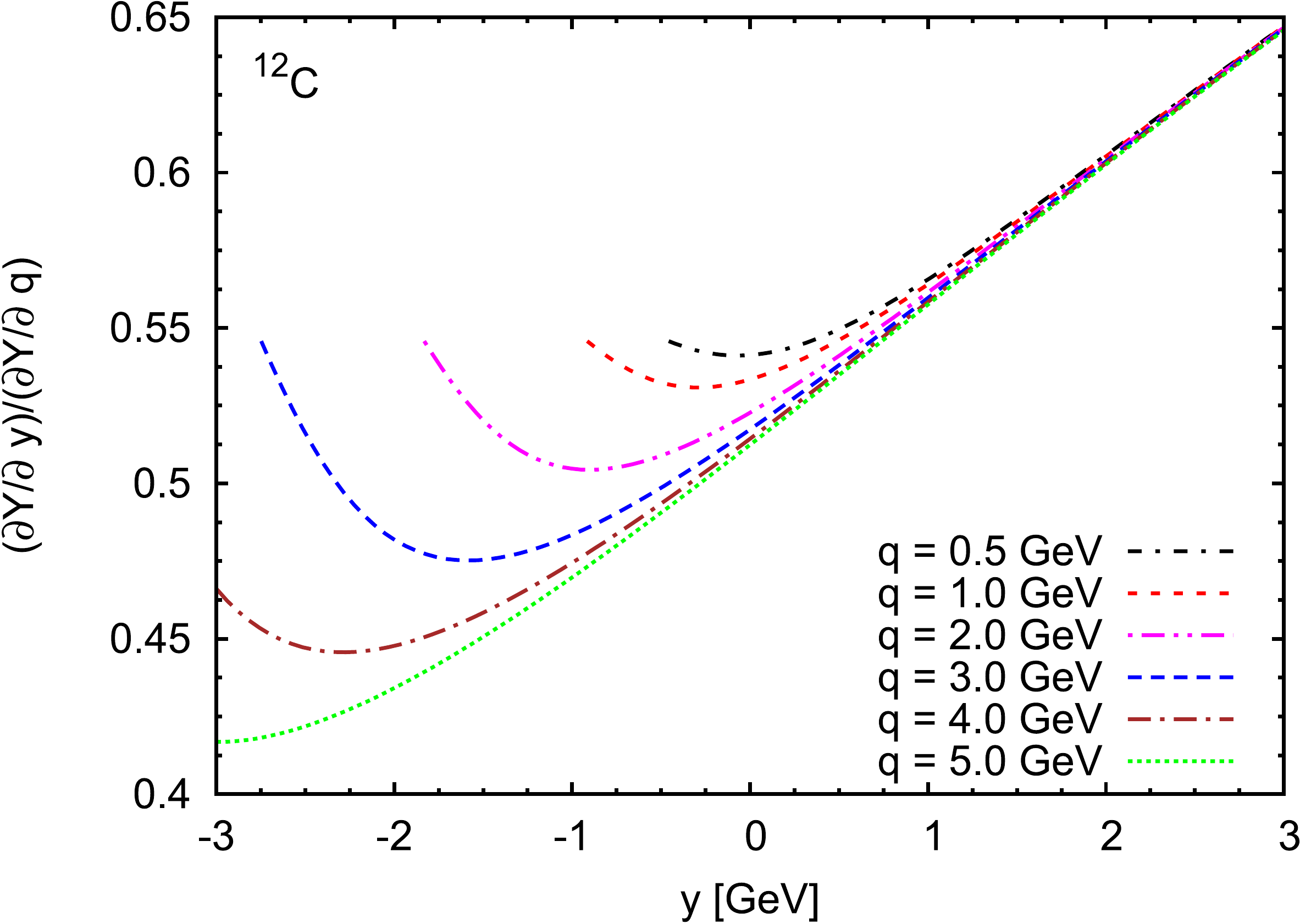}
\caption{(Color online) Results for the $(\partial Y/\partial y)/(\partial Y/\partial q)$ for $^{12}$C nucleus using analytical expressions of derivatives obtained from Eq.~(\ref{eq:Ylarge_app}).\label{dYdy_dYdq12C}}
\end{minipage}\hfill
\begin{minipage}[t]{0.49\textwidth}
\centering\includegraphics[width=\textwidth]{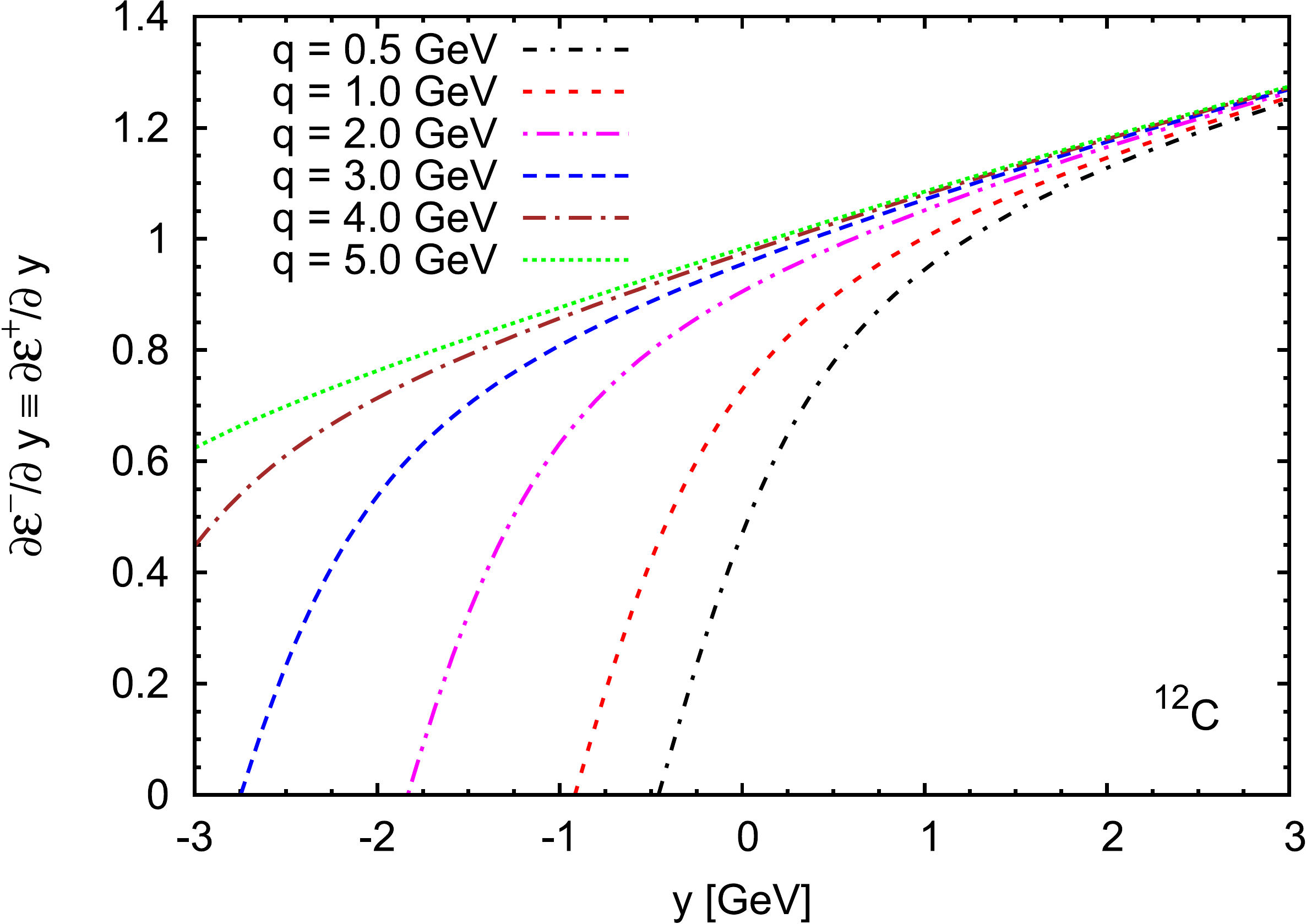}
\caption{(Color online) Results for the $\partial {\cal E}^\pm/\partial y$ for $^{12}$C nucleus obtained from Eq.~(\ref{eq:exc1}).
\label{fig:depsdyC12}}
\end{minipage}
\end{figure*}

In Fig.~\ref{fig:diffY_C12} are presented the derivatives $\partial Y(q,y)/\partial y$ [panel (a)] and $\partial Y(q,y)/\partial q$ [panel (b)] at fixed values of $q$. The results are obtained using analytical expressions of derivatives obtained from Eq.~(\ref{eq:Ylarge_app}). The black solid line, shown as reference, refers to the corresponding approximate expressions of the derivatives in the thermodynamic limit $M^0_B\rightarrow\infty$ ($\partial Y/\partial y \simeq 1$ and $\partial Y/\partial q \simeq 2$). As can be seen from the figures, in the case of light nuclei, as the $^{12}$C nucleus considered, the use of the exact expression for the derivatives leads to results that deviate very significantly from the thermodynamic limit. In Fig.~\ref{dYdy_dYdq12C} we present the ratio of the derivatives which is a part of Eqs.~(\ref{eq:D1}) and~(\ref{eq:D2}). The behaviour of the ratio at positive $y$ is almost the same for all $q$-values considered. In Fig.~\ref{fig:depsdyC12} are shown the derivatives of the excitation energy (\ref{eq:exc1}) with respect to $y$:
\begin{equation}
  \frac{\partial {\cal E}^-}{\partial y}=\frac{\partial {\cal E}^+}{\partial y}=
\frac{y}{\sqrt{{M_B^0}^2 + y^2}} + \frac{q+y}{\sqrt{m_N^2 + (q + y)^2}}.
\end{equation}

\begin{figure*}[htb]
\centering
\includegraphics[width=.495\textwidth]{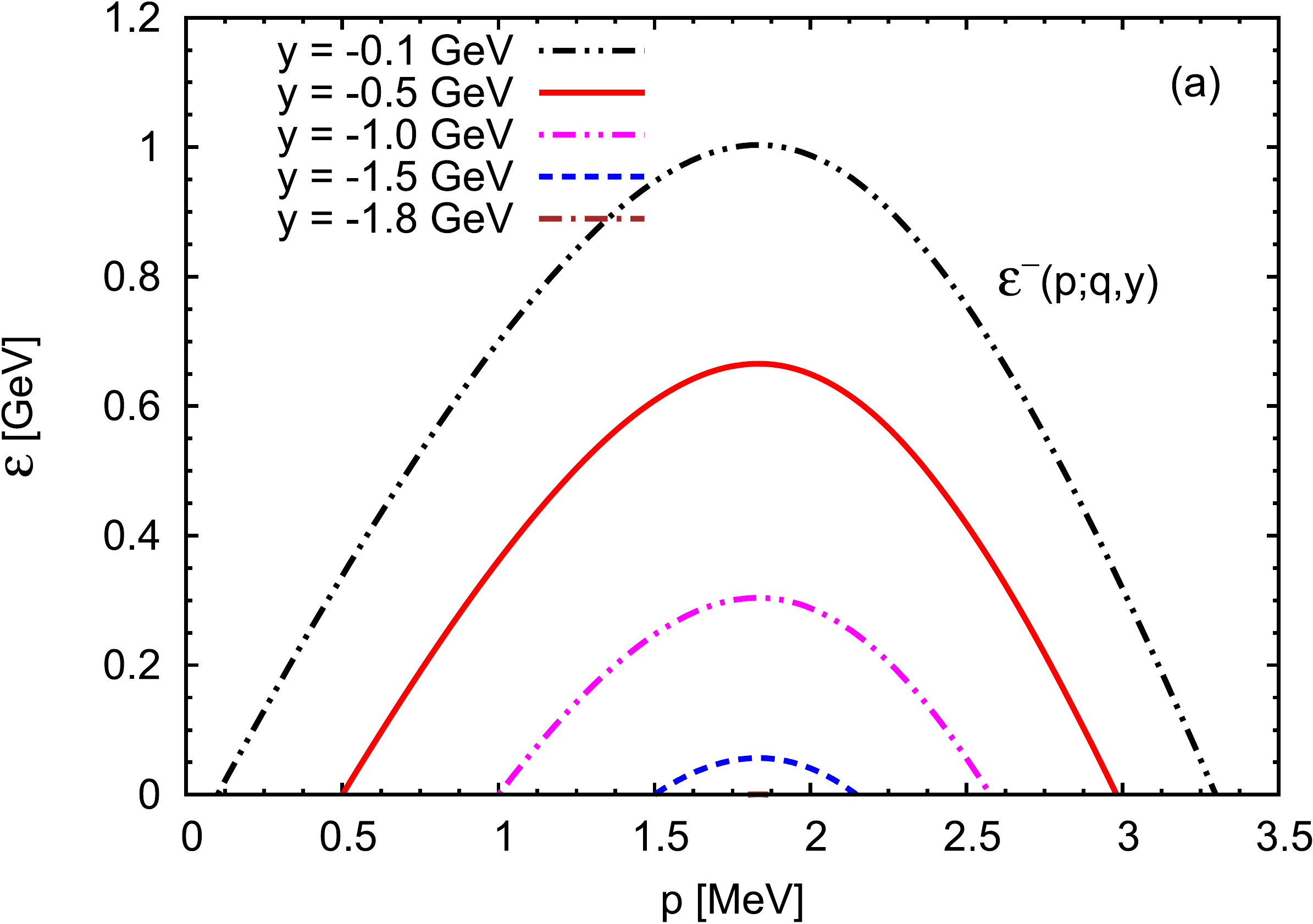}\hfill\includegraphics[width=.495\textwidth]{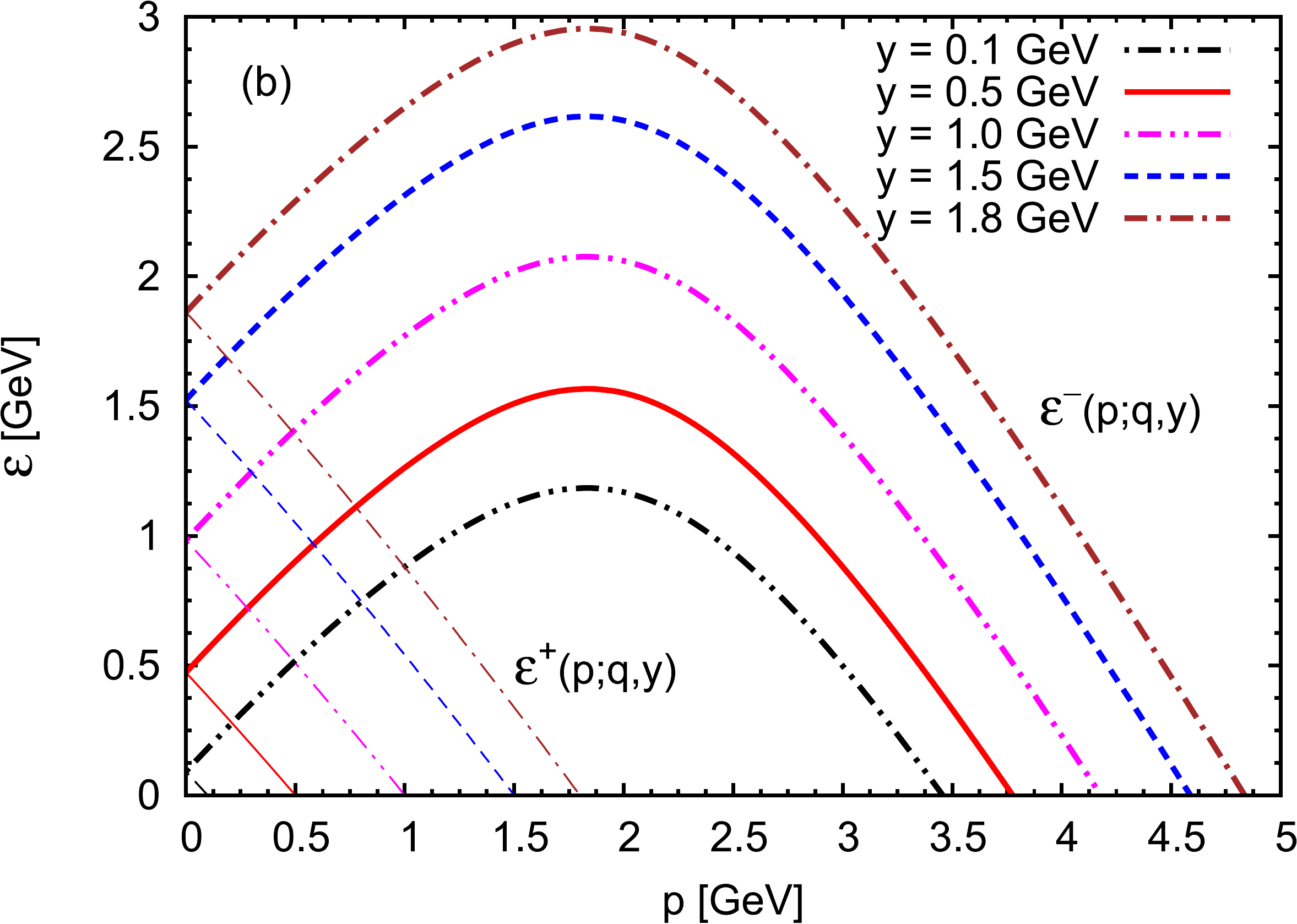}
\caption{(Color online) Excitation energy ${\cal E}$ corresponding to negative (left panel) and positive (right panel) values of $y$ at fixed momentum transfer $q=2$~GeV. \label{fig:eps_12C}}
\end{figure*}

\begin{figure*}[htb]
\begin{minipage}[t]{0.49\textwidth}
\centering\includegraphics[width=\textwidth]{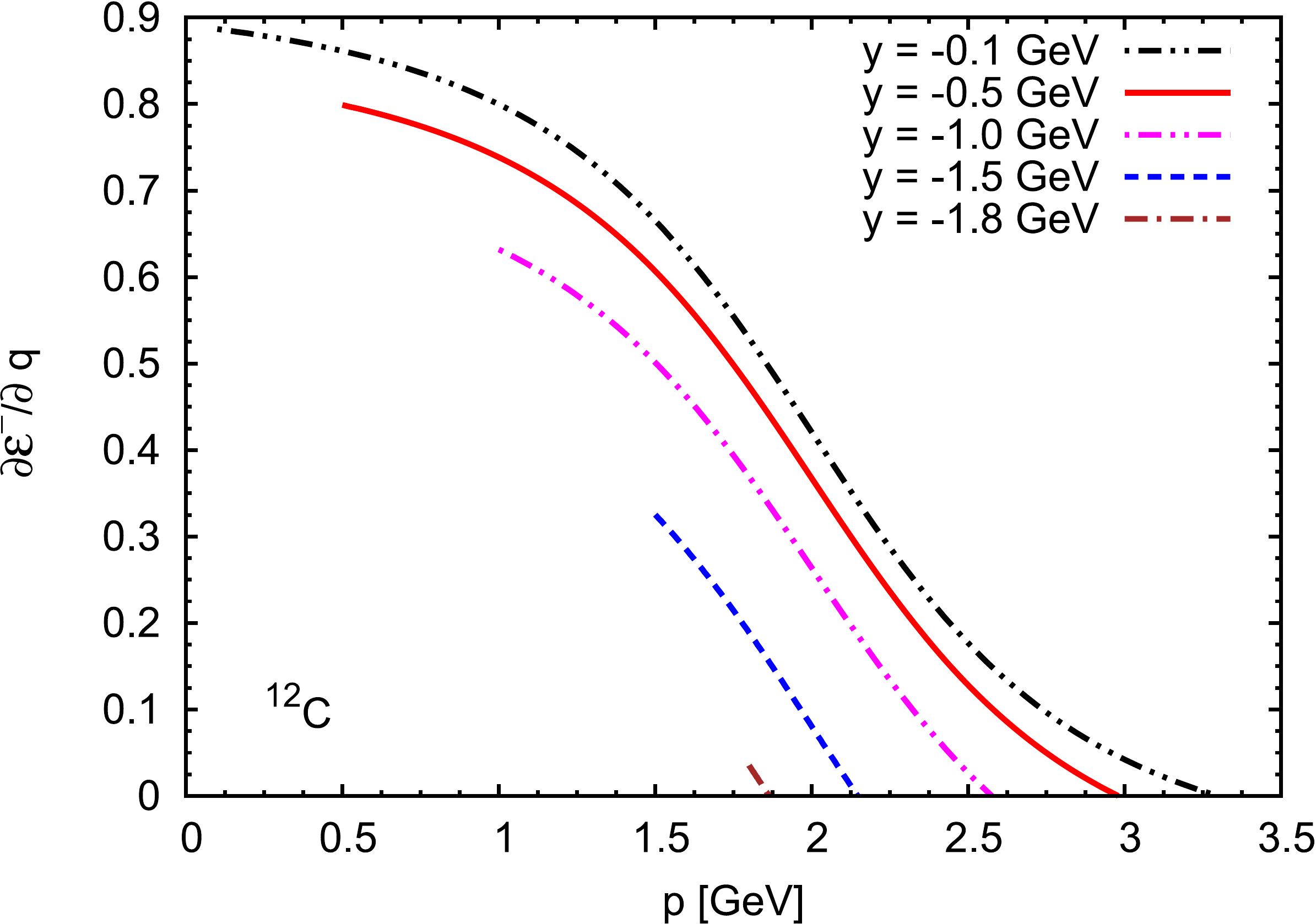}
\caption{(Color online) Results for the $\partial {\cal E}^{-}/\partial q$ corresponding to negative values of $y$ at fixed
momentum transfer $q = 2$~GeV.\label{fig:depsdq_12C_neg}}
\end{minipage}\hfill
\begin{minipage}[t]{0.49\textwidth}
\centering\includegraphics[width=\textwidth]{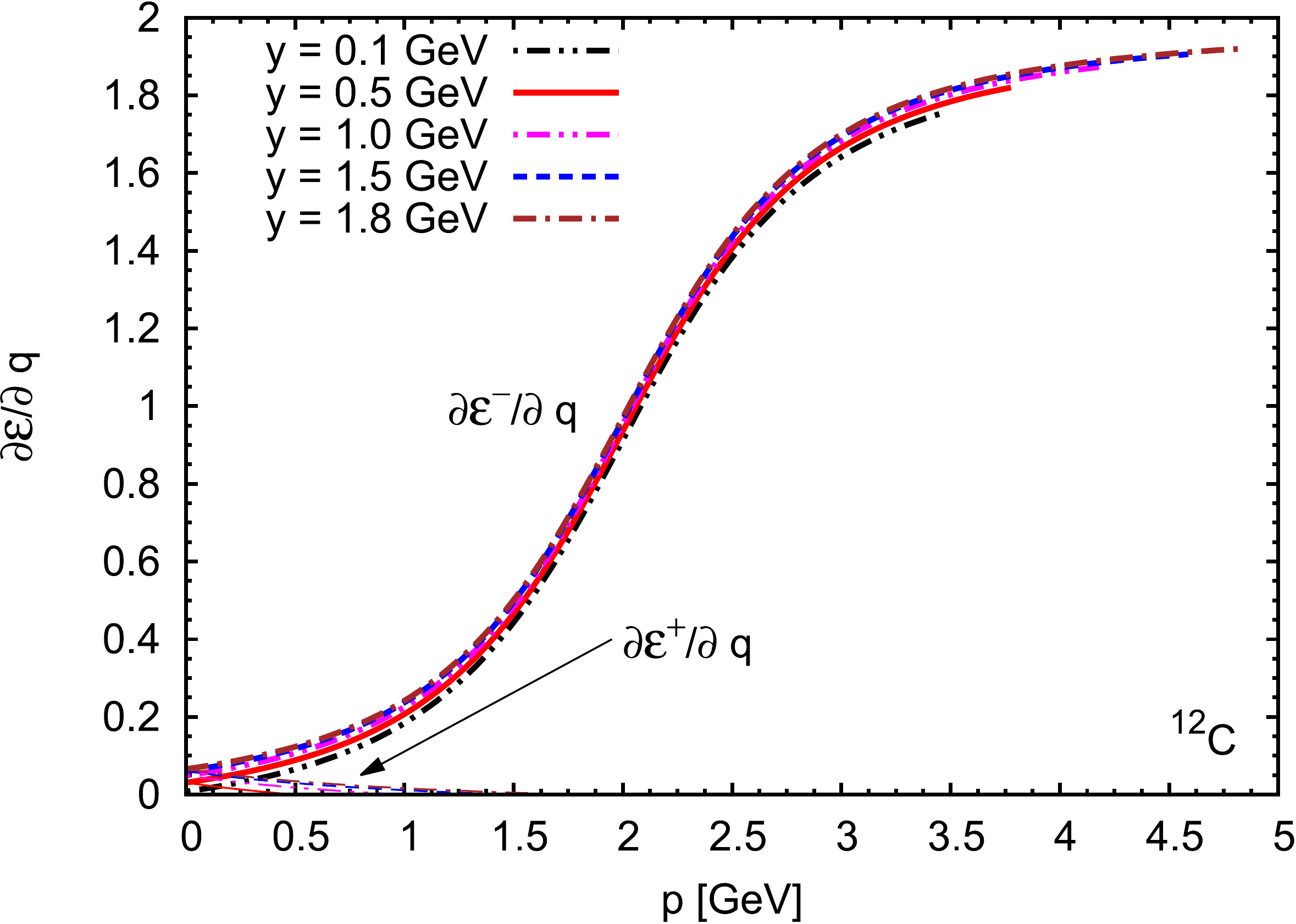}
\caption{(Color online) Results for the $\partial {\cal E}/\partial q$ corresponding to positive values of $y$ at fixed momentum transfer $q = 2$~GeV. \label{fig:depsdq_12C_pos}}
\end{minipage}
\end{figure*}

In Figs.~\ref{fig:eps_12C}--\ref{fig:depsdq_12C_pos} are presented results for the excitation energy and derivatives of the excitation energy with respect to $q$. In these figures we fix the momentum transfer $q=2$~GeV, and assume scaling of the first kind to be fulfilled [see Eq.~(\ref{eq:sc.f.I_kind}) and~\ref{appendixC}]. The allowed region of integrations [$-y \leq p \leq Y(q,y)$, $0\leq {\cal E}\leq {\cal E}^-(p;q;y)$] given in Eq.~(\ref{eq:Fneg}) is shown in Fig.~\ref{fig:eps_12C} (left panel) for several fixed negative values of $y$. The right panel shows the allowed region of integrations [$0\leq p \leq y$, ${\cal E}^+(p;q;y) \leq {\cal E}\leq {\cal E}^-(p;q;y)$ and $y\leq p \leq Y(q,y)$, $0\leq {\cal E}\leq {\cal E}^-(p;q;y)$] given in Eq.~(\ref{eq:Fpos}) for several fixed positive values of $y$, where ${\cal E}^+(p;q;y)$ and ${\cal E}^-(p;q;y)$ are drawn by thin and thick lines, respectively. For completeness, in Figs.~\ref{fig:depsdq_12C_neg} and~\ref{fig:depsdq_12C_pos} are shown the derivatives of the excitation energy ${\cal E}^{\pm}(p;q,y)$ [Eq.~(\ref{eq:exc1})] with respect to $q$ at negative and positive values of $y$, respectively.

\subsection{Superscaling Function \emph{vs.} Nucleon Momentum Distribution\label{sec:superscaling}}

The superscaling variable $\psi$ is introduced by (see Refs.~\cite{PhysRevC.60.065502, PhysRevLett.82.3212, Barbaro1998137, PhysRevC.38.1801}):
\be
\psi =\frac{1}{\sqrt{\xi_F}} \frac{\lambda - \tau}{\sqrt{(1+\lambda) \tau + \kappa \sqrt{\tau (1+\tau)}}} \, , \label{psi-RFG}
\ee
where $\lambda \equiv \omega / 2 m_N$, $\kappa \equiv q/ 2 m_N$ and $\tau\equiv |Q^2|/ 4 m_N^2  = \kappa^2 -\lambda^2$. The scaling variables $y$ and $\psi$ are closely connected:
\be
\psi = \left(\frac{y}{k_F} \right) \left[ 1 +\sqrt{1+\frac{m_N^2}{q^2}}\frac{1}{2} \eta_F \left( \frac{y}{k_F} \right) +{\cal O} [\eta_F^2] \right]\,,\label{eq:y_psi}
\ee
where $\eta_F\equiv k_F/m_N$ and, as noted above, the superscaling function $f$ is connected with $F$ via $f\equiv k_F\times F$ with
$k_F$ the Fermi momentum. Then we can write:
\be
\dfrac{\partial F}{\partial y}= \dfrac{1}{k_F}\cdot\dfrac{\partial f(\psi)}{\partial \psi}\cdot\dfrac{\partial \psi}{\partial y}=
\dfrac{1}{k_F^2}\cdot\dfrac{\partial f(\psi)}{\partial \psi}\cdot {\cal V}(y) ,\label{eq:dFdpsi}
\ee
where
\be
{\cal V}(y)= 1 +\sqrt{1+\frac{m_N^2}{q^2}} \eta_F \left( \frac{y}{k_F}\right) \, .
\ee
The Eq.~(\ref{eq:n0_neg}) for negative values of $y$ ($y=-k$) can be written as:
\be
\left(\frac{\partial f(\psi)}{\partial \psi}\right)_{y=-k} =A_1(y)+A_2(y)=A(y), \label{eq:sc.f.vs.y_neg}
\ee
where
\be
A_1(y)= -2\pi k_F^2 \Bigg[ \dfrac{y\, n_0(-y)}{{\cal V} (y)} \Bigg]_{y=-k} \,,\quad
A_2(y)= {2\pi k_F^2} \Bigg[ \int\limits_{-y}^{Y(q,y)} \dfrac {{\cal D}_1(p;q,y)\, S_1(p,\,{\cal E}^-)} {{\cal V} (y)}
\,p\,dp  \Bigg]_{y=-k}. \label{eq:sc.f.vs.y_neg_cont}
\ee

The corresponding result for positive values of $y$ [Eq.~(\ref{eq:n0_pos})] can be written as:
\be
\left(\frac{\partial f(\psi)}{\partial \psi}\right)_{y=k} =B_1(y)+B_2(y)+B_3(y)=B(y), \label{eq:sc.f.vs.y_pos}
\ee
with
\begin{gather}
B_1(y)= -2\pi k_F^2 \Bigg[ \dfrac{y\, n_0(y)}{{\cal V} (y)} \Bigg]_{y=k}\,,\quad
B_2(y)= {2\pi k_F^2} \Bigg[ \int\limits_{0}^{Y(q,y)} \dfrac {{\cal D}_1(p;q,y)\, S_1(p,\,{\cal E}^-)} {{\cal V} (y)}
\,p\,dp  \Bigg]_{y=k}\,,\notag\\
B_3(y)= -2\pi k_F^2  \Bigg[ \int\limits_{0}^{y} \dfrac {{\cal D}_2(p;q,y)\, S_1(p,\,{\cal E}^+)} {{\cal V} (y)} \,p\,dp \Bigg]_{y=k}. \label{eq:sc.f.vs.y_pos_cont}
\end{gather}

Finally, comparing Eq.~(\ref{eq:sc.f.vs.y_neg}) and Eq.~(\ref{eq:sc.f.vs.y_pos}), we can write the following connection between the superscaling function $f(\psi)$ and $S_1(p,{\cal E})$
\begin{multline}
\mathfrak{F}(k)=\Bigg(\frac{\partial f(\psi)}{\partial \psi}\cdot{\cal V}(y)\Bigg)_{y=-k}
+ \Bigg(\frac{\partial f(\psi)}{\partial \psi}\cdot{\cal V}(y)\Bigg)_{y=k} ={2\pi k_F^2} \Bigg[ \int\limits_{k}^{Y(q,-k)}
\!\!\! p dp \left[ {{\cal D}_1(p;q,y)\, S_1(p,\,{\cal E}^-)} \right]_{y=-k}+\\
+\!\!\!\int\limits_0^{Y(q,k)}\!\!\! p\,dp\, \left[ {{\cal D}_1(p;q,y)\, S_1(p,\,{\cal E}^-)} \right]_{y=k} -\int\limits_0^k p\,dp\,\left[{{\cal D}_2(p;q,y)\, S_1(p,\,{\cal E}^+)}\right]_{y=k} \Bigg].\label{eq:sc.f.vs.SF}
\end{multline}
Here, we note that Eq.~(\ref{eq:sc.f.vs.SF}) is obtained using a general expression of the scaling function given in terms of the nuclear spectral function within PWIA [see Eq.~(\ref{eq:scaling_function})] and split into two terms, corresponding to zero and finite excitation energy [see Eq.~(\ref{eq:spectral})]. From Eq.~(\ref{eq:sc.f.vs.SF}) is clearly visible the connection between the first derivative of the superscaling function and the term corresponding to the finite excitation energy $S_1(p,\,{\cal E})$ of the spectral function. In the case of using the approximate expression of Eq.~(\ref{eq:y_psi}) ($\psi\simeq y/k_F$)  we can write:
\begin{equation}
\mathfrak{F}(k)=\Bigg(\frac{\partial f(\psi)}{\partial \psi}\Bigg)_{\psi=-\psi_0} + \Bigg(\frac{\partial f(\psi)}{\partial \psi}\Bigg)_{\psi=\psi_0},
\end{equation}
where $\psi_0 = k/ k_F$ and therefore the scaling function is symmetric [$f(\psi) = f(-\psi)$] when $S_1(p,{\cal E})=0$.

In the general case the scaling function $F(q,\omega)$ is related to the spectral function $S(p,{\cal E})$ by Eq.~(\ref{eq:scaling_function}). After some mathematical manipulation and using the assumption that scaling of the first kind is fulfilled, we find the connection between $n(k)$ and $f(\psi)$ as illustrated by Eqs.~(\ref{eq:sc.f.vs.y_neg})--(\ref{eq:sc.f.vs.SF}). In our approach we intend to extract information about the nucleon momentum distribution (which is the function of variable $k$) from the experimental superscaling function (which is the function of variable $\psi$). It is important to note that the connection [Eq.~(\ref{eq:sc.f.vs.SF})] is similar to the one shown by C. Ciofi degli Atti \emph{et al.}, but here we extend our analysis of the behaviour of the superscaling function not only to the case of negative $\psi$ but also to that of positive $\psi$.

On this point we would like to note the earlier works on the problem of the experimental determination of the nucleon momentum distribution, \emph{e.g.} those of Frankel Ref.~\cite{PhysRevLett.38.1338, PhysRevC.17.694} and Amado and Woloshin Refs.~\cite{PhysRevLett.36.1435, AMADO1977400}. It has been shown in Ref.~\cite{AMADO1977400} that the final state interactions (FSI) destroy the simple dependence of the inclusive cross section [$d^3\sigma/d^3q$, where $q$ is the momentum of the observed proton after collision] on the momentum distribution $n(k)$. However, it has been pointed out in Ref.~\cite{PhysRevLett.38.1338, PhysRevC.17.694} a way ``to retain the benefit of scaling  by replacing $n(k)$ by an \emph{effective momentum distribution} $n_\text{eff}(k)$''. A procedure has been developed that relates the differential cross section to the ground state wave function and to the FSI both of which had to come from the solution of the appropriate many-body problem with the true Hamiltonian. Thus, it has been concluded in Ref.~\cite{PhysRevLett.38.1338, PhysRevC.17.694} that the fact that the cross section cannot be related directly to the ground state momentum distribution $n(k)$, as shown in Ref.~\cite{AMADO1977400}, is ``not a real loss''.  It turned out that the effective momentum distribution is (roughly) proportional to the actual momentum distribution $n(k)$. For instance, it has been pointed out in Ref.~\cite{PhysRevLett.38.1338, PhysRevC.17.694} that if $n(k)$ decreases exponentially with $k$, then $n_\text{eff}(k)$, that incorporates FSI, also decreases exponentially with $k$.

\section{Results\label{sec:results}}

\begin{figure}[htb]
\centering\includegraphics[width=.7\textwidth]{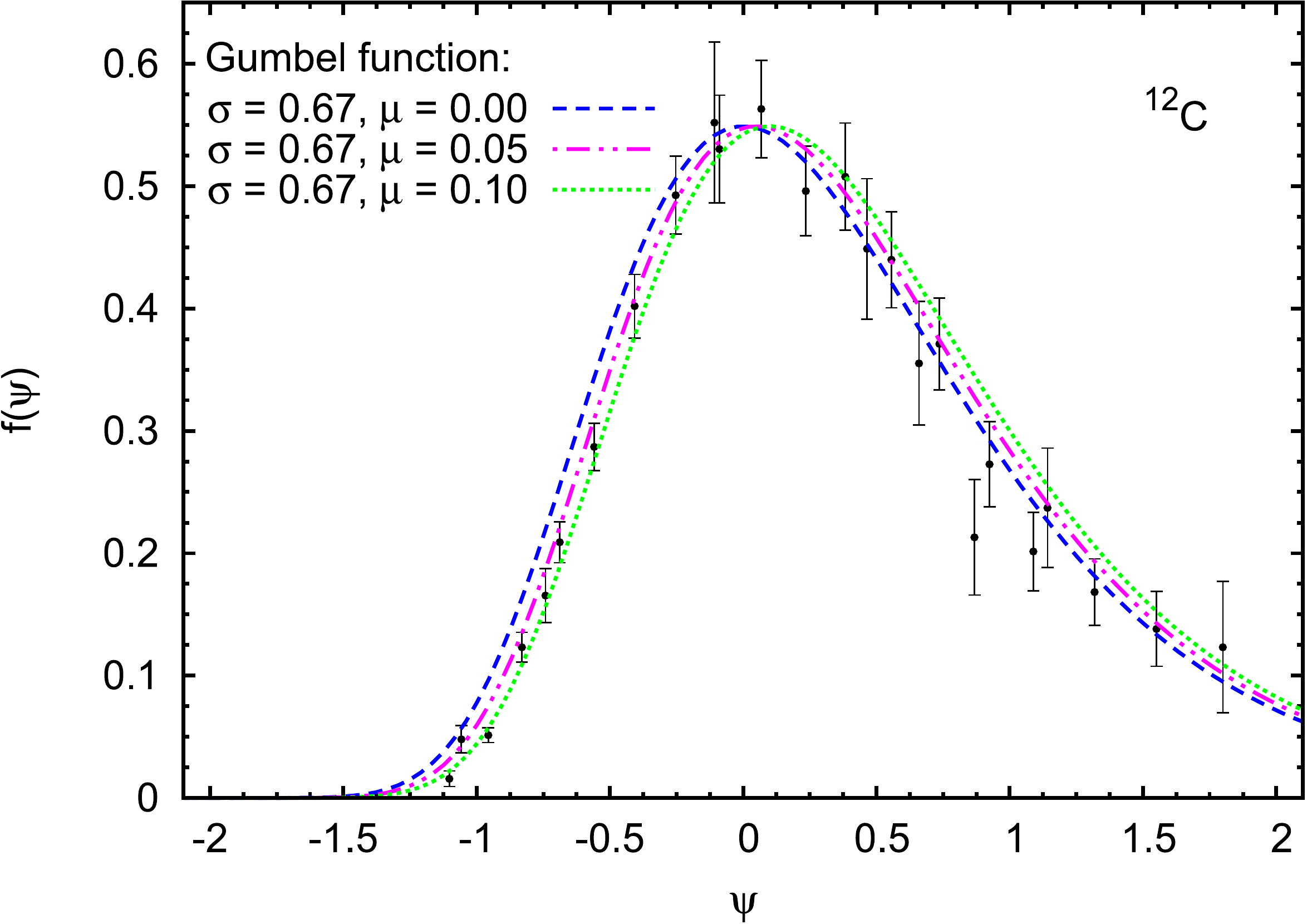}
\caption{(Color online) Averaged experimental $f(\psi)$ versus $\psi$ in the quasielastic region together with a phenomenological parametrization of the data using Gumbel distribution function~Eq.~(\ref{eg:Gumbel}) with three sets of parameters ($\sigma=0.67$, $\mu=0.00$; $\sigma=0.67$, $\mu=0.05$; $\sigma=0.67$, $\mu=0.10$). The integral of the curve has been normalized to unity. \label{fig:fpsi}}
\end{figure}

\begin{figure}[htb]\centering
\centering\includegraphics[width=.7\textwidth]{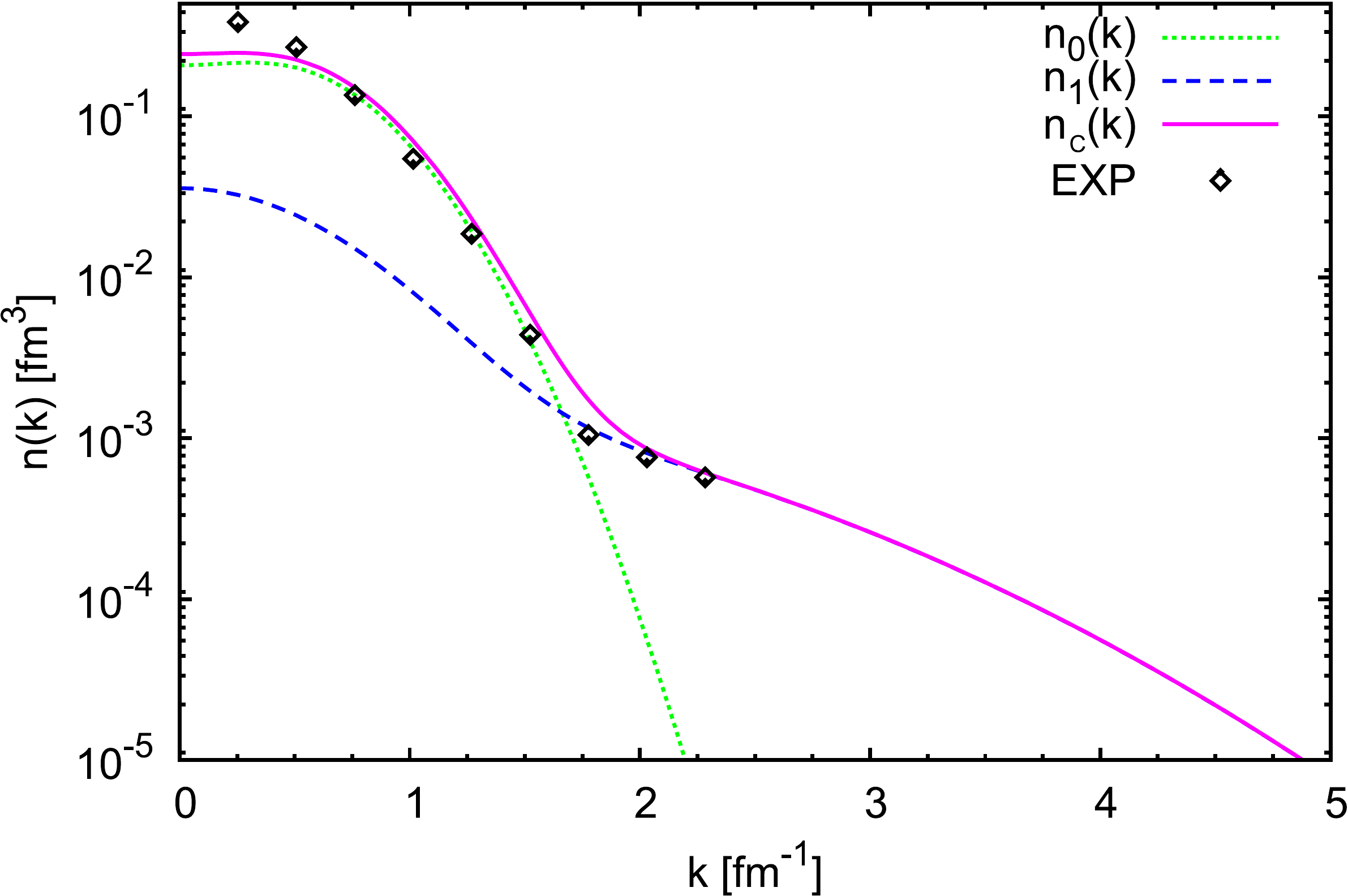}
\caption{The many-body nucleon momentum distribution $n_C(k)$ corresponding to the parametrization described in the Appendix of Ref.~\cite{PhysRevC.53.1689}. The normalization of $n(k)$ is given in Eq.~(\ref{eq:norm.n(k)}).\label{fig:nk12C}}
\end{figure}

In this Section we present our analysis of the nucleon momentum distribution extracted from the experimental scaling function. At sufficiently high energies are seen both types of scaling behavior (see Ref.~\cite{PhysRevC.71.015501} and references therein). For specific nuclei one observes quite good first-kind scaling at excitation energies below the QE peak, namely, in the so-called scaling region. This is the familiar $y$-scaling behavior. On the other hand, it is known from the available data where longitudinal-transverse separations have been made, that these scaling violations apparently reside in the transverse response, but not in the longitudinal. The latter appears to superscale. In fact, this is not unexpected, since there are contributions that do not scale arising from meson-exchange currents (MECs) plus the correlation effects required by gauge invariance which must be considered together with the MEC~\cite{AMARO2002317, AMARO2002388, AMARO2003181}, and from inelastic scattering from the nucleons~\cite{Barbaro:2003ie}. It is important to note that MEC and inelastic contributions are predominantly transverse in the kinematic regions of interest in the present work. In Fig.~\ref{fig:fpsi} are presented the averaged longitudinal experimental data of the superscaling function $f(\psi)$ versus $\psi$ in the quasielastic region together with a phenomenological parametrization of the data using a Gumbel distribution function. The Gumbel distribution function is defined as
\be
f_G(\psi)=\dfrac{1}{\sigma} \exp\left[-\dfrac{(\psi-\mu)}{\sigma}\right] \exp\left[-\exp\left[-\dfrac{(\psi-\mu)}{\sigma}\right]\right], \label{eg:Gumbel}
\ee
where $\sigma$ and $\mu$ are scale and location parameters, respectively. We use three sets of parameters of the Gumbel distribution function: the scale parameter $\sigma$ is fixed to $0.67$ and three values for $\mu=0.00,~ 0.05,~0.10$. The integral of the Gumbel function has been normalized to unity.

In the present work we consider the spectral function in the form given by Eq.~(\ref{eq:spectral}). For the nucleon momentum distribution $n(k)$ for $^{12}$C we make use of the parametrization described in the Appendix of Ref.~\cite{PhysRevC.53.1689}, see Fig.~\ref{fig:nk12C} (in what follows noted by $n_\text{C}(k)$). The momentum density $n(k)$ is normalized to $1$, \textit{i.e.},
\be
4\pi\int\limits_0^{\infty}n(k)k^2dk=1\,,\label{eq:norm.n(k)}
\ee
where $n(k)=n_0(k)+n_1(k)$. Interacting nucleons (beyond the mean field approach) are described by the correlated part of the spectral function $S_1(p,{\cal E})$. As known (see Ref.~\cite{PhysRevC.41.R2474} and references therein), the two-nucleon interactions dominate. The short-range correlations give rise to pairs of nucleons with high relative momentum. We follow the approach of Kulagin and Petti~\cite{KULAGIN2006126}, where it is assumed (see also Ref.~\cite{PhysRevC.53.1689}) that $S_1(p,{\cal E})$ at high momentum and high separation energy is dominated by ground state configurations with a correlated nucleon-nucleon pair and the remaining $(A - 2)$ nucleons moving with low center-of-mass momentum. In this approach interactions of higher order are not included. Then, $S_1(p,{\cal E})$ can be expressed analytically in the form (see also~\cite{PhysRevC.77.044311}):
\be
S_1(p,{\cal E})=n_1(p)\,\frac{M}{p}\,\sqrt{\frac\alpha\pi}\left(e^{-\alpha\, p_{\text{min}}^2}-e^{-\alpha\, p_{\text{max}}^2}\right)\,,\label{eq:correlationSF}
\ee
where
$$\alpha=\dfrac{3}{4\langle  p_0^2\rangle \beta}\,,\quad \beta=\frac{A-2}{A-1}\,.$$
The mean square of the $n_0(p)$ momentum $\langle  p_0^2\rangle$ is defined as
\begin{equation}\label{eq:<p^2_0>}
\langle p_0^2\rangle=\dfrac{4\pi\int p^4\, n_0(p)\, d p}{4\pi\int p^2\, n_0(p)\, dp}\,,
\end{equation}
whereas
\begin{equation}\label{eq:pmin,pmax}
p_{\text{min/max}}^2=\Big\{\beta\,  p \mp \sqrt{2\,M\,\beta\,{\cal E}}\,\Big\}^2\,.
\end{equation}
The excitation energy, given as the difference between the missing and separation energies, is given by
\be
{\cal E} = E_{\text{m}}- E^{(2)}_S\,,\label{eq:ciofi.vs.our1}
\ee
where the two-nucleon separation energy $E^{(2)}_S$ is an average excitation of the $(A-2)$ nucleon system. Since by definition averaging should be carried out only over the low-lying states, it can be approximated by the mass difference $E^{(2)}_S=M_{B}^0+2\,m_N-M_A^0\cong M_{A-2}+2\,m_N-M_A$.

\begin{figure}[htb]
\centering\includegraphics[width=.7\textwidth]{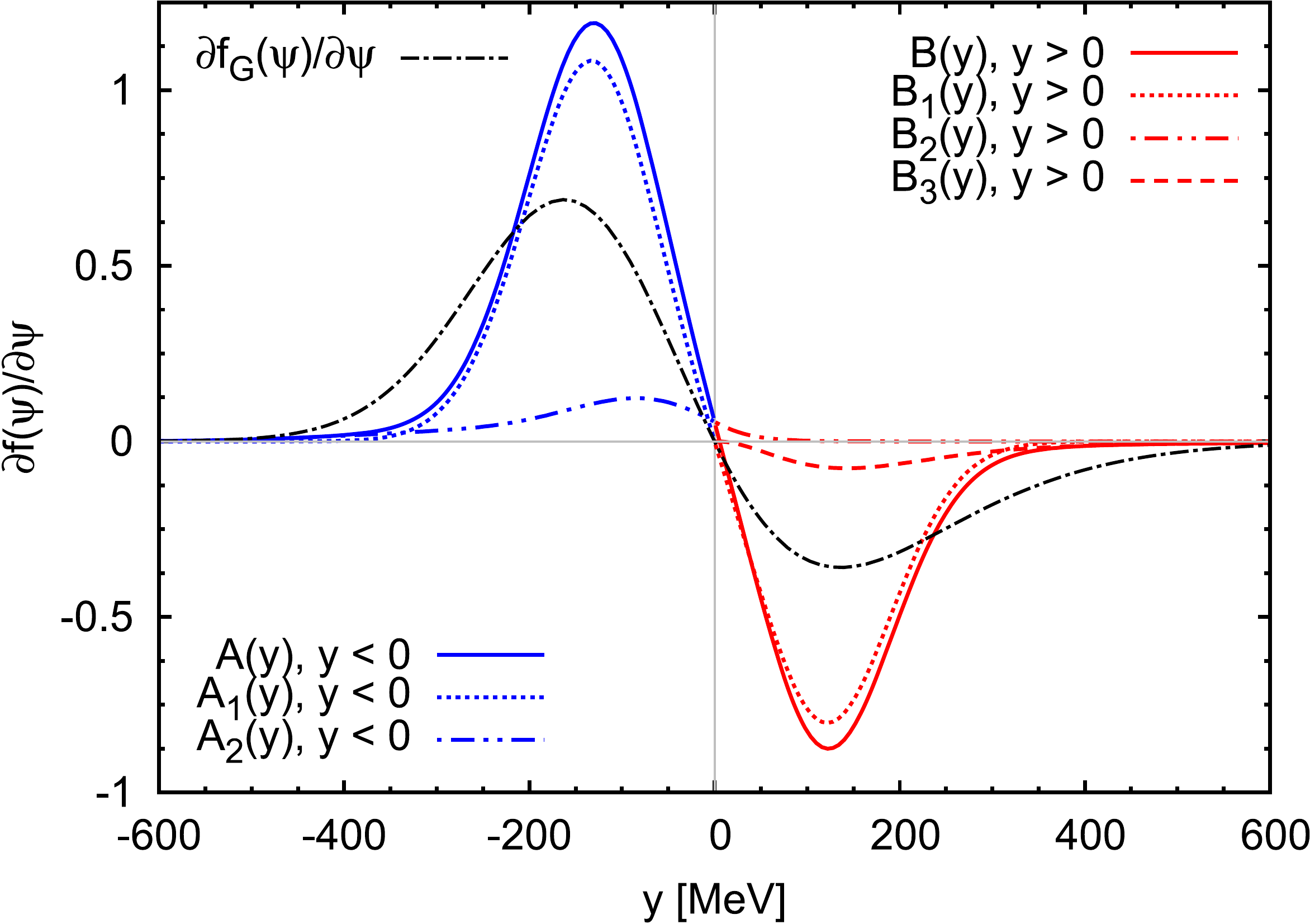}
\caption{(Color online) Results for the first derivative of the scaling function ${\partial_{\psi} f(\psi)}$ for negative [Eq.~(\ref{eq:sc.f.vs.y_neg})] and positive [Eq.~(\ref{eq:sc.f.vs.y_pos})] values of $y$. For more details, see the text.\label{fig:dfdpsi.vs.y}}
\end{figure}

\begin{figure}[htb]
\centering\includegraphics[width=.8\textwidth]{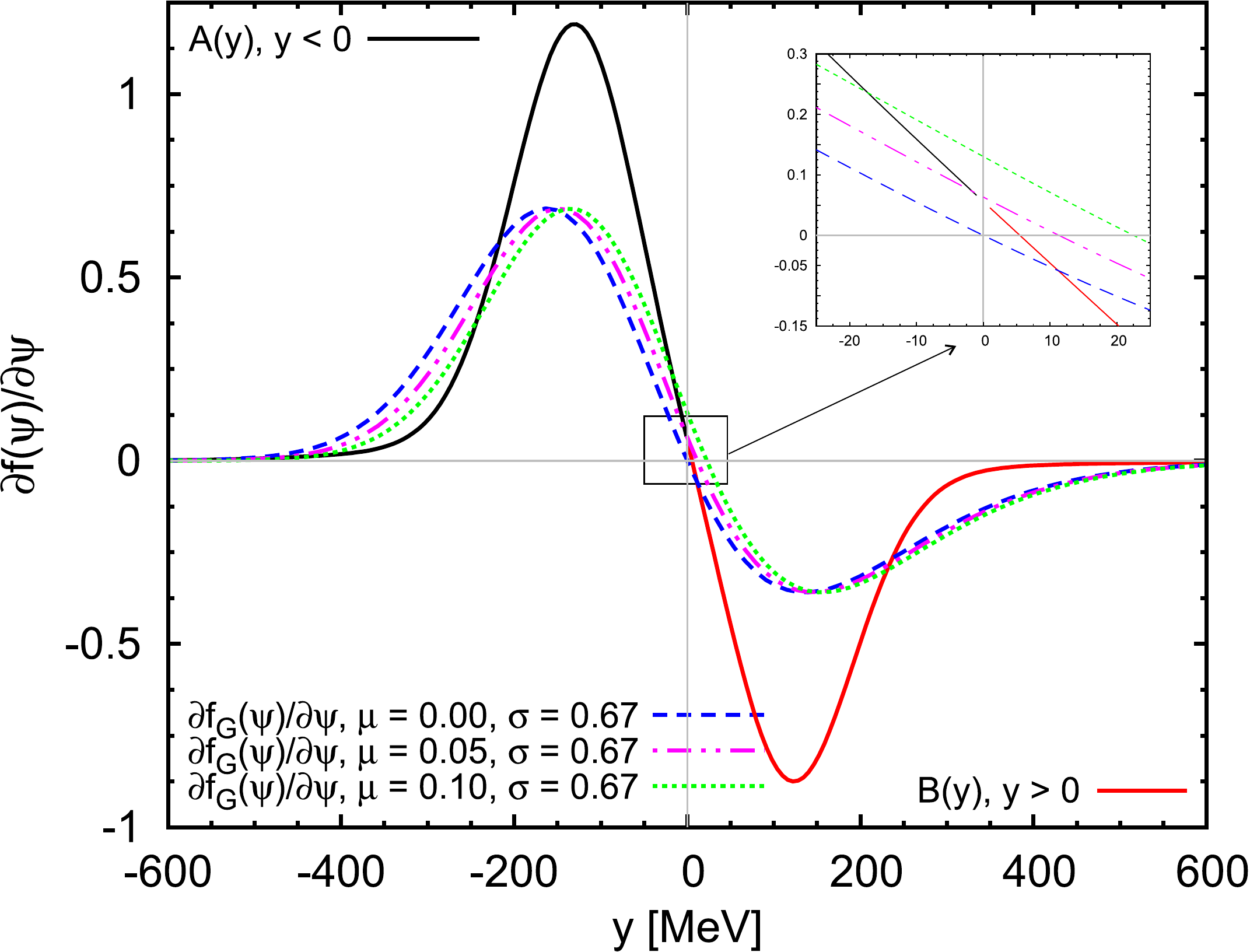}
\caption{(Color online) Results for the first derivative of the scaling function ${\partial_{\psi} f_G(\psi)}$ using the three sets of parameters of the Gumbel distribution function in comparison with $A(y)$ and $B(y)$ [see Eqs.~(\ref{eq:sc.f.vs.y_neg}) and ~(\ref{eq:sc.f.vs.y_pos})]. \label{fig:dfdpsi.vs.y1}}
\end{figure}

\begin{figure}[htb]
\centering\includegraphics[width=.8\textwidth]{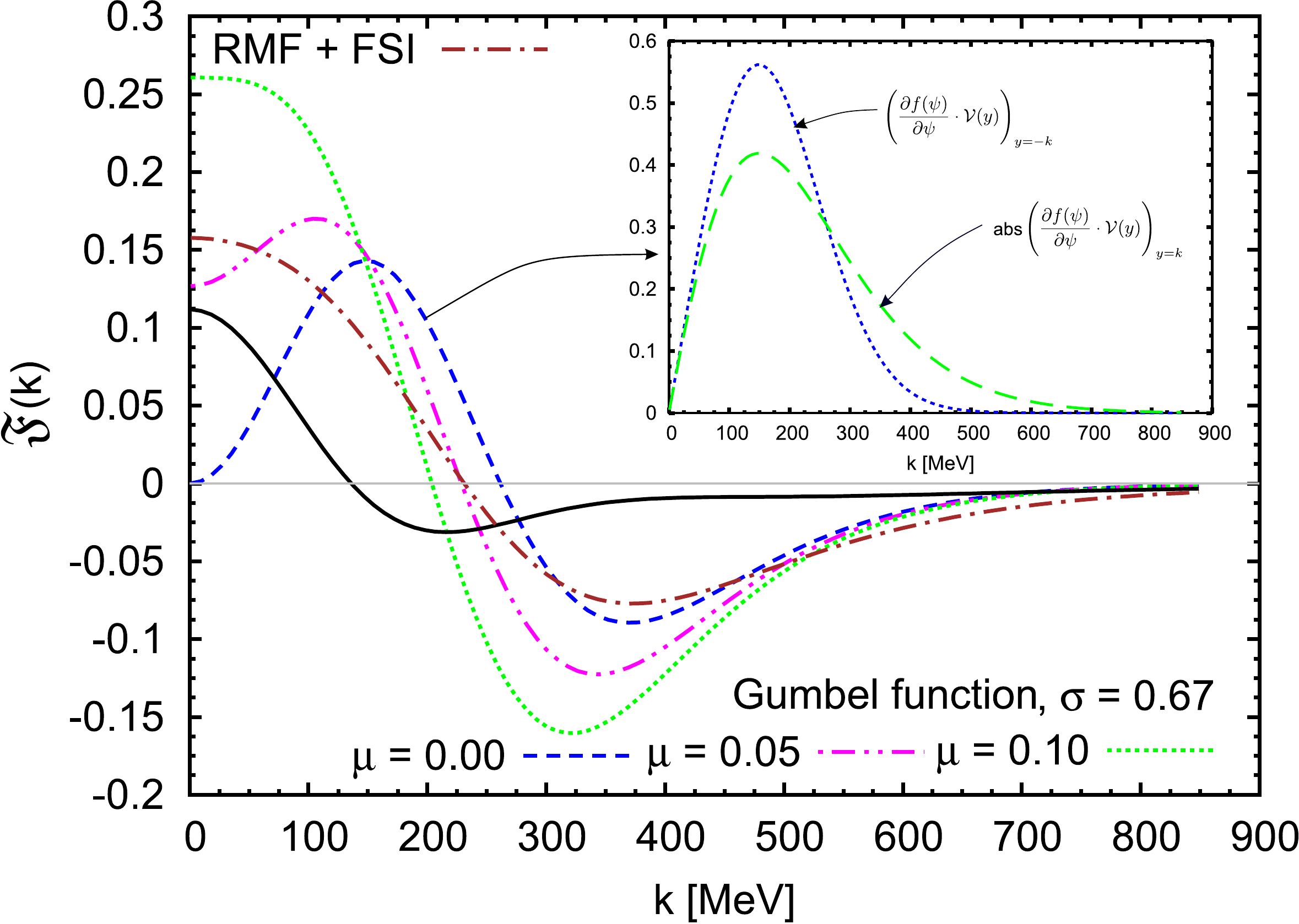}
\caption{(Color online) Results for $\mathfrak{F}(k)$ function using the three sets of parameters of the Gumbel distribution function [left-hand side of Eq.~(\ref{eq:sc.f.vs.SF})] in comparison with the predictions obtained by the right-hand side of Eq.~(\ref{eq:sc.f.vs.SF}) using spectral function [see Eqs.~(\ref{eq:spectral}) and~(\ref{eq:correlationSF})] and momentum distributions: $n_\text{C}(k)$ described in the Appendix of Ref.~\cite{PhysRevC.53.1689} (black solid line) and  RMF + FSI given in the~\ref{appendixB} (brown dash-dotted line).\label{fig:FF}}
\end{figure}

\begin{figure}[htb]
\centering\includegraphics[width=\textwidth]{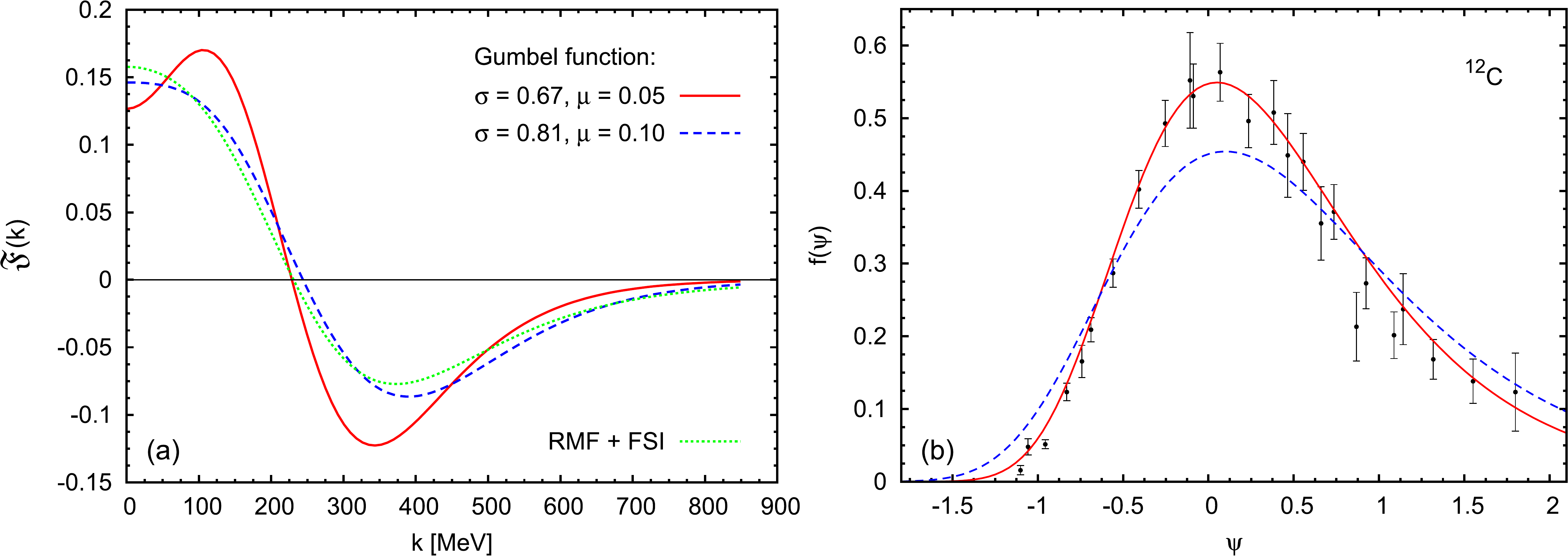}
\caption{(Color online) Results for $\mathfrak{F}(k)$ function [panel (a)] using two sets (see text) of parameters of the Gumbel distribution function [left-hand side of Eq.~(\ref{eq:sc.f.vs.SF})] in comparison with the predictions obtained by the right-hand side of Eq.~(\ref{eq:sc.f.vs.SF}) using spectral function and momentum distribution within RMF + FSI approach (green dotted line). $f(\psi)$ versus $\psi$ [panel (b)] in the quasielastic region together with Gumbel distribution function~Eq.~(\ref{eg:Gumbel}) with two sets of parameters ($\sigma=0.67$, $\mu=0.05$ and $\sigma=0.81$, $\mu=0.10$).\label{fig:FF1}}
\end{figure}

In Fig.~\ref{fig:dfdpsi.vs.y} we present results for the first derivative of the scaling function ${\partial_{\psi} f(\psi)}$ for negative [Eq.~(\ref{eq:sc.f.vs.y_neg})] and positive [Eq.~(\ref{eq:sc.f.vs.y_pos})] values of $y$. The black dash-dotted line represents the first derivative of the Gumbel distribution function using $\sigma = 0.67$ and $\mu=0.0$. The blue solid line corresponds to using the right-hand side of Eq.~(\ref{eq:sc.f.vs.y_neg}), with the separate contributions $A_1(y)$ (blue dotted line) and $A_2(y)$ (blue dot-dot-dashed line). The red solid line displays the result using the right-hand side of Eq.~(\ref{eq:sc.f.vs.y_pos}) and corresponding contributions $B_1(y)$ (red dotted line), $B_2(y)$ (red dot-dot-dashed line), and $B_3(y)$ (red dashed line). Results are obtained using the spectral function in the form given by Eq.~(\ref{eq:spectral}) and momentum distributions [$n_0(k)$ and $n_1(k)$] taken from the Appendix of Ref.~\cite{PhysRevC.53.1689}. It is clearly visible that the main contribution comes from $A_1$ and $B_1$ terms, which are closely related to $n_0(k)$ momentum distribution. Notice that by using given parametrization and normalization of $n_0(k)$ it is not possible to describe the first derivative of the experimental scaling function ($A_1$ and $B_1$ overpredict derivative of the experimental scaling function).

Results for the first derivative of the scaling function ${\partial_{\psi} f_G(\psi)}$ using three sets of parameters of the Gumbel distribution function in comparison with $A(y)$ and $B(y)$ [see Eqs.~(\ref{eq:sc.f.vs.y_neg}) and~(\ref{eq:sc.f.vs.y_pos})] are presented in Fig.~\ref{fig:dfdpsi.vs.y1}. The inset depicts the behaviour of ${\partial_{\psi} f(\psi)}$ in the $ y \approx 0$~MeV region, which explains the different behaviour of the function $\mathfrak{F}(k)$ at $k=0$~MeV. In Fig.~\ref{fig:FF} are presented results for $\mathfrak{F}(k)$ using the three sets of parameters of the Gumbel distribution function [left-hand side of Eq.~(\ref{eq:sc.f.vs.SF})]. In this figure we show the sensitivity of the results to the parameters which are used to describe the experimental scaling function. Obviously, having experimental data with small error bars will allow us to extract more correct information about the scaling function and respectively on the nucleon momentum distribution. These results are compared with the predictions obtained by the right-hand side of Eq.~(\ref{eq:sc.f.vs.SF}) using spectral function [see Eqs.~(\ref{eq:spectral}) and~(\ref{eq:correlationSF})] and momentum distributions: $n_\text{C}(k)$ described in the Appendix of Ref.~\cite{PhysRevC.53.1689} (black solid line) and  from RMF + FSI given in the~\ref{appendixB} (brown dash-dotted line). As shown in Fig.~\ref{fig:FF} the result using spectral function and $n_\text{C}(k)$ does not provide a proper description of the $\mathfrak{F}(k)$ behaviour obtained from the analysis of the experimental scaling function (Gumbel distribution function). The agreement improves when using the RMF + FSI momentum distribution taken from Ref.~\cite{PhysRevC.83.045504} (our analytical fit to the RMF + FSI momentum distribution is given in~\ref{appendixB}): the minimum of $\mathfrak{F}(k)$ is between $300$ and $400$~MeV as in the case of the experimental Gumbel function, also the tail of $\mathfrak{F}(k)$ slightly overpredicts results when the Gumbel function is used. As shown in Ref.~\cite{PhysRevC.83.045504}, the RMF + FSI model leads to a scaling function $f (\psi)$ that, for positive values of $\psi$, is in good accordance with electron scattering data.

In Fig.~\ref{fig:FF1} is given the $\mathfrak{F}(k)$ function [panel (a)] obtained using two sets of parameters of the Gumbel distribution function [left-hand side of Eq.~(\ref{eq:sc.f.vs.SF})] in comparison with the predictions obtained by the right-hand side of Eq.~(\ref{eq:sc.f.vs.SF}) using spectral function and momentum distribution within RMF + FSI approach. The results present our attempt to find (as an example) sets of parameters of the Gumbel distribution function that give the best fit to the $\mathfrak{F}(k)$ function within RMF + FSI approach. One can see the curves corresponding to two sets of parameters: i) $\sigma=0.67$ and $\mu=0.05$ from the variation of $\sigma$ between $0.6$ and $0.8$ and of $\mu$ between $-0.1$ and $0.1$, and ii) $\sigma = 0.81$ and $\mu =0.1$ from the variation of $\sigma$ between $0.5$ and $1.0$ and of $\mu$ between $-0.2$ and $0.2$. The corresponding scaling functions $f(\psi)$ are shown in Fig.~\ref{fig:FF1} [panel (b)]. Although the second fit gives better description of $\mathfrak{F}(k)$ within RMF + FSI approach, it is still not possible to describe more correctly the experimental data of the scaling function. This shows the necessity to use a self-consistent procedure to search simultaneously for both the nucleon momentum distribution $n(k)$ and the scaling function $f(\psi)$.

Here it is important to point out that the analytical fit given in~\ref{appendixB} is not unique, because it is possible to use different forms of the parametrization of the $n_0(k)$ and $n_1(k)$ and therefore different normalization of the two parts of momentum distribution. Although Eq.~(\ref{eq:sc.f.vs.SF}) gives a direct connection between $S_1(p,{\cal E})$ and the experimental scaling function, the key point is to look for such a part $n_1(k)$ of the momentum distribution to be consistent with $n_0(k)$ and with the general normalization condition of the momentum distribution [Eq.~(\ref{eq:norm.n(k)})]. We are presently working on a self-consistent procedure to determine $n_0(k)$ and $n_1(k)$ that are consistent with the correct behaviour of $\mathfrak{F}(k)$ using experimental scaling function. The results will be presented in a forthcoming publication.

To conclude we would like to point out that the main contribution to the tail of $\mathfrak{F}(k)$ at high momentum, $k \ge 400$~MeV comes from positive $y$-values (likewise positive-$\psi$), as can be seen from the inset in Fig.~\ref{fig:FF}. The contributions $\text{abs} [{\partial_{\psi} f(\psi)}\cdot{\cal V}(y)]$ to the $\mathfrak{F}(k)$ function from positive and negative values of $y$ obtained using the Gumbel distribution function ($\sigma =0.67 $, $\mu = 0 $) are shown in the inset in Fig.~\ref{fig:FF}.

\section{Conclusions\label{sec:conclusions}}

Electron scattering has been considered over years to be one of the most powerful tools to get precise information on the structure of nuclei. In particular, the analysis of semi-inclusive $(e,e'p)$ and inclusive $(e,e')$ reactions have unambiguously proved the shell structure in nuclei, and have also provided detailed knowledge on nucleon-nucleon correlations. A key point to consider in this area concerns the connection between the observables extracted directly from the experiment and those theoretical concepts related to nuclear properties. This question has not an easy response because the general description of the reaction mechanism requires to have a good control over very different ingredients. The basic objective of this work is focused on the link between the scaling function, an observable extracted directly from the analysis of $(e,e')$ data, and the nuclear spectral function and/or the momentum density, that refers directly to the inner structure of the nucleus. This is a complex problem that has been treated in previous works by different groups. In this sense we mentioned in particular studies using the Green's function method and its representation of the spectral function. This technique has been applied to nuclear matter and finite nuclei providing results in good agreement with data extracted from electron scattering experiments, and clarifying the role of short and long-range correlations in different experimental quantities.

In what follows we emphasize the importance of scaling ideas. As known, this behavior emerges from the analysis of inclusive QE $(e,e')$ data. The scaling function does contain information on how the nucleons are distributed in the nuclear target, but also on different ingredients that play a key role in the reaction mechanism. Final state interactions between the ejected nucleon and the residual nucleus in addition to effects beyond the impulse approximation, {\it i.e.,} meson exchange currents, and even higher nucleon inelasticities are captured by the scaling function extracted from the analysis of data. This explains the interest of scaling arguments to be connected with more or less sophisticated nuclear models to be used in scattering reaction studies. Notice that the electromagnetic responses can be explored using a variety of models, but the scaling function sets a strong constraint to any model aimed at describing lepton-nucleus scattering processes.

In this work, starting with a general expression of the nuclear spectral function split into two terms corresponding to zero and finite excitation energy, we have developed the explicit equations that connect the spectral function (or momentum distribution) with the derivatives of the scaling function. We take into account the dependence of the scaling function in $q$ and $y$ (transfer momentum and scaling variable), and present an analysis that incorporates the regions below and above the QE peak (negative and positive $y$-values, respectively). A very detailed study on the behavior of the different derivatives involved in the process is presented.

Using the Gumbel distribution density to describe the superscaling function extracted from the analysis of the separate longitudinal $(e,e')$ data, and different theoretical approximations to deal with the short-range correlations in the spectral function, the present work contains novel results dealing with the close link between both magnitudes. We have adopted the notation introduced originally by C. degli Atti and collaborators, but have extended their conclusions by maintaining the full $(q,y)$-dependence in the scaling function and the whole positive and negative $y$-region in the analysis. A more systematic and self-consistent procedure to determine the global momentum distribution in accordance with the experimental scaling function is in progress and results will be presented in a forthcoming publication.

\section*{Acknowledgements}

This work was partially supported by the Bulgarian National Science Fund under contract No. KP-06-N38/1, by the Spanish Ministerio de Economia y Competitividad and ERDF (European Regional Development Fund) under contracts FIS2017-88410-P, by the Junta de Andalucia (FQM 160, SOMM17/6105/UGR) and by the Spanish Consolider-Ingenio 2000 program CPAN (CSD2007-0042),

\appendix

\section{\label{appendixA}}

In this Appendix we show how Eqs.~(\ref{eq:n0_neg}) and~(\ref{eq:n0_pos}) are obtained from Eqs.~(\ref{eq:Fneg1}) and~(\ref{eq:Fpos1}), respectively.

\subsection{Negative--$y$ region}

\begin{gather}
{\dfrac{1}{2\pi}}\frac{\partial F}{\partial y} = Y\,n_0(Y)\left(\frac{\partial Y}{\partial y}\right)-y\,n_0(-y)+ \int\limits_{-y}^{Y(q,y)} p\,dp\,\left(\frac{\partial{\cal E}^-}{\partial y}\right)S_1(p,\,{\cal E}^-)\,, \label{eq:dFdyneg}\\
{\dfrac{1}{2\pi}}\frac{\partial F}{\partial q} = Yn_0(Y)\left(\frac{\partial Y}{\partial q}\right)+\int\limits_{-y}^{Y(q,y)} p\,dp\,\left(\frac{\partial {\cal E}^-}{\partial q}\right) S_1(p,\,{\cal E}^-) \, .\label{eq:dFdqneg}
\end{gather}
Multiplying Eq.~(\ref{eq:dFdyneg}) by $(\partial Y/\partial q)$ and Eq.~(\ref{eq:dFdqneg}) by $(\partial Y/\partial y)$ and subtracting the two equations, we get:
\be
{\dfrac{1}{2\pi}}\frac{\partial F}{\partial y}\left(\frac{\partial Y}{\partial q}\right)-
{\dfrac{1}{2\pi}}\frac{\partial F}{\partial q}\left(\frac{\partial Y}{\partial y}\right)
=-y\,n_0(-y)\left(\frac{\partial Y}{\partial q}\right)+\int\limits_{-y}^{Y(q,y)} p\,dp\,
\left(\frac{\partial{\cal E}^-}{\partial y}\frac{\partial Y}{\partial q}-\frac{\partial {\cal E}^-}{\partial q}\frac{\partial Y}{\partial y}\right)S_1(p,\,{\cal E}^-)\, .\label{eq:diff_Fneg}
\ee
Making use the limit of Eq.~(\ref{eq:sc.f.I_kind}) in Eq.~(\ref{eq:diff_Fneg}) and $y=-k$, we simply have
\be
n_0(k)= {\dfrac{1}{2\pi k}}\left(\frac{\partial F}{\partial y}\right)_{y=-k} - \frac{1}{k}\left[\int\limits_{-y}^{Y(q,y)} \mathcal{D}_1(p;q,y)\, S_1(p,\,{\cal E}^-)\,p\,dp \right]_{y=-k}.\label{eq:diff_Fneg1}
\ee

\subsection{Positive--$y$ region}

\begin{gather}
{\dfrac{1}{2\pi}}\frac{\partial F}{\partial y} = Y\,n_0(Y)
\left(\frac{\partial Y}{\partial y}\right)-y\,n_0(y)+\int\limits_0^{Y(q,y)} p\,dp\,S_1(p,\,{\cal E}^-)
\left(\frac{\partial {\cal E}^-}{\partial y}\right)-\int\limits_0^y p\,dp\,S_1(p,\,{\cal E}^+)
\left(\frac{\partial {\cal E}^+}{\partial y}\right)\label{eq:dFdypos}\\
{\dfrac{1}{2\pi}}\frac{\partial F}{\partial q} = Yn_0(Y)\left(\frac{\partial Y}{\partial q}\right)
+\int\limits_0^{Y(q,y)} p\,dp\,S_1(p,\,{\cal E}^-) \left(\frac{\partial {\cal E}^-}{\partial q}\right) -
\int\limits_0^y p\,dp\,S_1(p,\,{\cal E}^+)\left(\frac{\partial {\cal E}^+}{\partial q}\right) \, .\label{eq:dFdqpos}
\end{gather}
Multiplying Eq.~(\ref{eq:dFdypos}) by $(\partial Y/\partial q)$ and Eq.~(\ref{eq:dFdqpos}) by $(\partial Y/\partial y)$ and subtracting the two equations, we get:
\begin{multline}
{\dfrac{1}{2\pi}}\frac{\partial F}{\partial q}\left(\frac{\partial Y}{\partial y}\right) - {\dfrac{1}{2\pi}}\frac{\partial F}{\partial y}\left(\frac{\partial Y}{\partial q}\right) = y\,n_0(y)\left(\frac{\partial Y}{\partial  q}\right)+\int\limits_0^{Y(q,y)} p\,dp\, \left(\frac{\partial {\cal E}^-}{\partial q}\frac{\partial Y}{\partial y}
-\frac{\partial {\cal E}^-}{\partial y}\frac{\partial Y}{\partial q} \right)S_1(p,\,{\cal E}^-)  \\
-\int\limits_0^y p\,dp\, \left(\frac{\partial {\cal E}^+}{\partial q}\frac{\partial Y}{\partial y}-
\frac{\partial {\cal E}^+}{\partial y}\frac{\partial Y}{\partial q} \right)S_1(p,\,{\cal E}^+)\label{eq:diff_Fpos}
\end{multline}
Making use the limit of Eq.~(\ref{eq:sc.f.I_kind}) in Eq.~(\ref{eq:diff_Fpos}) and $y=k$, we obtain
\begin{multline}
n_0(k) = - {\dfrac{1}{2\pi k}}\left(\frac{\partial F}{\partial y}\right)_{y=k} + \frac{1}{k}\left[\int\limits_0^{Y(q,y)} \mathcal{D}_1(p;q,y)\,S_1(p,\,{\cal E}^-)\,p\,dp\right]_{y=k}\\ - \frac{1}{k}\left[\int\limits_0^y  \mathcal{D}_2(p;q,y)\,S_1(p,\,{\cal E}^+)\,p\,dp\right]_{y=k}. \label{eq:diff_Fpos1}
\end{multline}

\section{\label{appendixB}}

\begin{figure}[h!]\centering
\centering\includegraphics[width=0.7\textwidth]{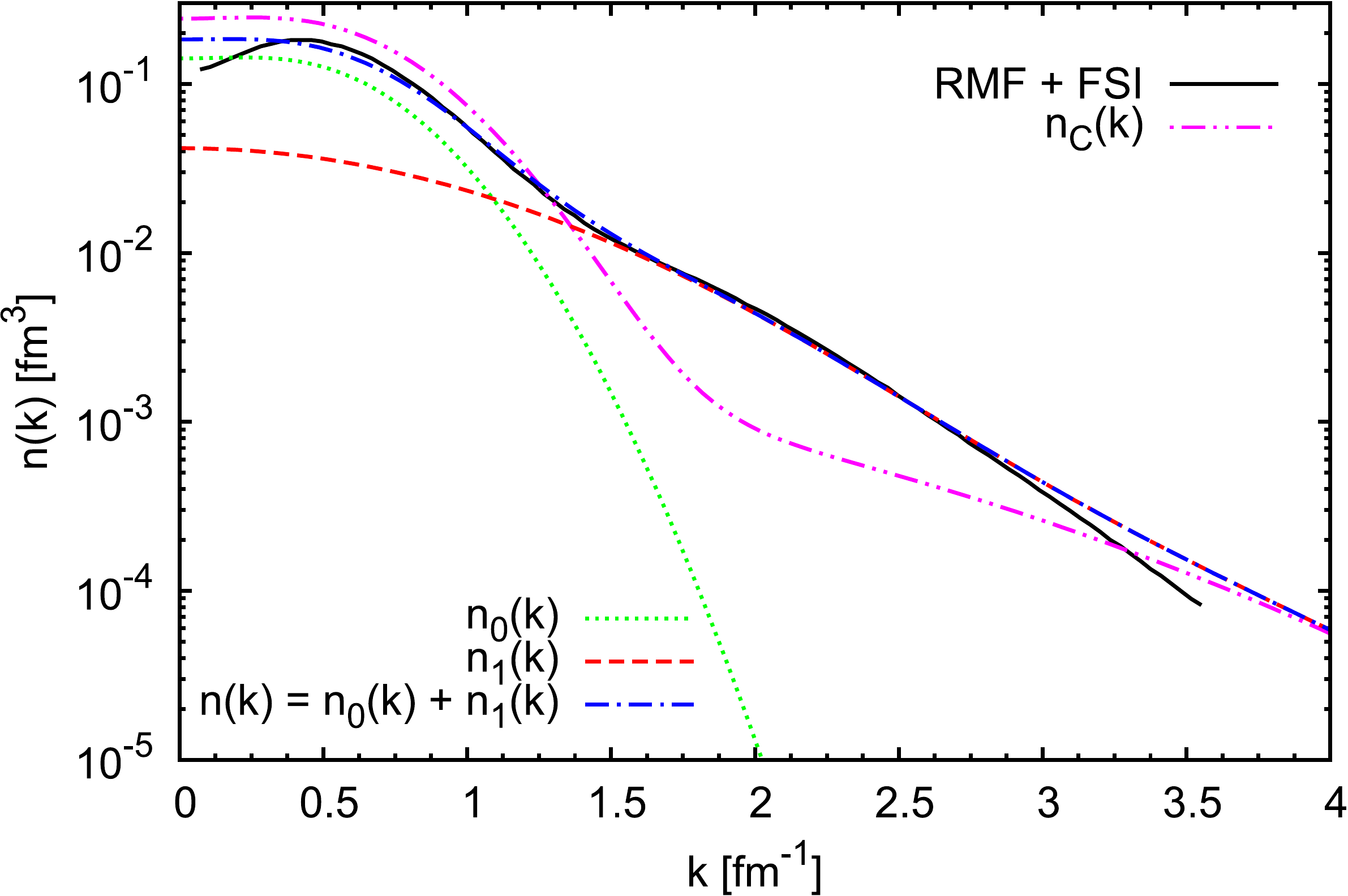}
\caption{The nucleon momentum distribution $n(k)$ corresponding to the parametrization described in the~\ref{appendixB} in comparison with $n_C(k)$. The normalization of $n(k)$ is given in Eq.~(\ref{eq:norm.n(k)}).\label{fig:nk12C_RMF}}
\end{figure}

In this Appendix we show some simple parametrizations of the nucleon momentum distributions obtained within the Relativistic Mean Field Approximation including Final State Interactions (denoted as RMF + FSI). The reader interested in more details can go to Ref.~\cite{PhysRevC.83.045504}. The expressions for the two contributions in the momentum density are written in the form:
$$n_0(k)={1.78\,e^{-3\,k^2}[1 + 3.54\,k^2]}/{(4\,\pi)}$$
and
$$n_1(k)=[0.5\,e^{-0.6\,k^2}+0.0237\,e^{-0.22\,k^2}]/(4\,\pi).$$
Fig.~\ref{fig:nk12C_RMF} shows the behavior of the two parametrized densities compared to the full RMF+FSI prediction. As known, the tail at high values of the momentum is entirely given by $n_1$. For completeness, the nucleon momentum distribution $n_C(k)$ corresponding to the parametrization described in the Appendix of Ref.~\cite{PhysRevC.53.1689} is also shown in Fig.~\ref{fig:nk12C_RMF}. As can be seen in Fig.~\ref{fig:nk12C_RMF} the behaviour of the two momentum distributions is quite different, that explains different results for $\mathfrak{F}(k)$ function shown in Fig.~\ref{fig:FF}. It is important to note that the nucleon momentum distribution obtained within RMF + FSI approach is an \emph{effective momentum distribution} because it is extracted using theoretically calculated inclusive electron cross section within RMF including FSI (for more details see Ref.~\cite{PhysRevC.83.045504}).

\section{\label{appendixC}}

\begin{figure}[h!]\centering
\centering\includegraphics[width=0.7\textwidth]{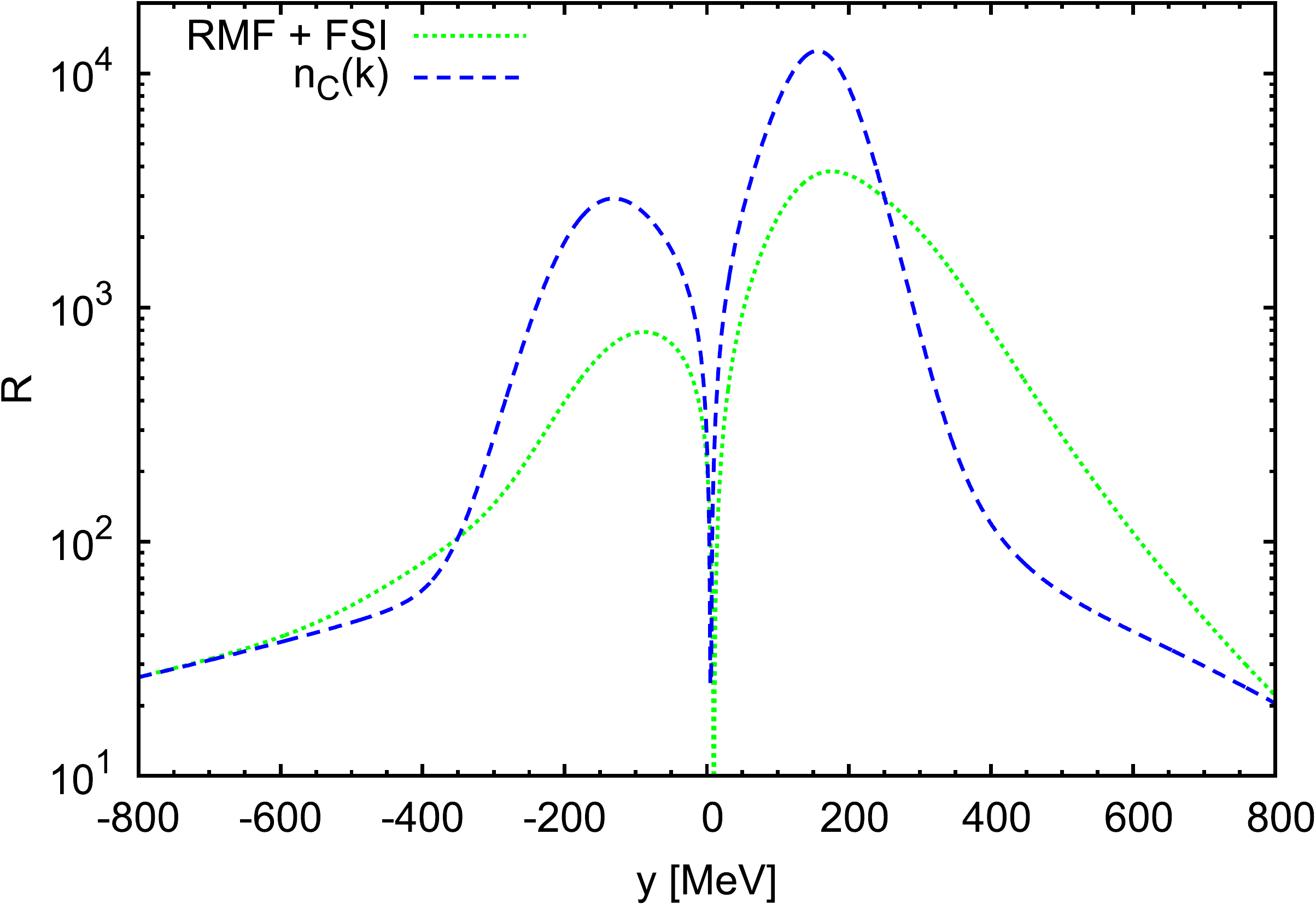}
\caption{Results for the ratio $R$ [Eq.~(\ref{eq:ratio})] using two different momentum distributions: RMF + FSI (\ref{appendixB}) and $n_\text{C}(k)$ (the parametrization described in the Appendix of Ref.~\cite{PhysRevC.53.1689}).\label{fig:ratio}}
\end{figure}

In this Appendix we check the validity of using the scaling of the first kind [Eq.~(\ref{eq:sc.f.I_kind})] in  Eqs.~(\ref{eq:diff_Fneg}) and~(\ref{eq:diff_Fpos}) for negative and positive values of $y$, respectively. For this purpose we compare the two terms from the left-hand side of Eqs.~(\ref{eq:diff_Fneg}) and~(\ref{eq:diff_Fpos}):
\be
\text{R}=\text{abs}\left[\left({\frac{\partial F}{\partial y}\cdot\frac{\partial Y}{\partial q}}\right) \Bigg/ \left({\frac{\partial F}{\partial q}\cdot\frac{\partial Y}{\partial y}}\right)\right]\label{eq:ratio}
\ee
using Eqs.~(\ref{eq:dFdyneg}) and~(\ref{eq:dFdqneg}) for negative values of $y$ and Eqs.~(\ref{eq:dFdypos}) and~(\ref{eq:dFdqpos}) for positive values of $y$, for two different momentum distributions: RMF + FSI (\ref{appendixB}) and $n_\text{C}(k)$ (the parametrization described in the Appendix of Ref.~\cite{PhysRevC.53.1689}). As can be seen in Fig.~\ref{fig:ratio}, the value of the ratio $R$ is larger than $20$ for the whole area of considered momentum $y$. Then the following relation can be placed:
\be
\Bigg|\!\frac{\partial F}{\partial y}\left(\frac{\partial Y}{\partial q}\right)\!\Bigg|>20\, \frac{\partial F}{\partial q}\left(\frac{\partial Y}{\partial y}\right)\,.\label{eq:ratio1}
\ee
The consequence of Eq.~(\ref{eq:ratio1}) is:
\be
{\dfrac{1}{2\pi}}\frac{\partial F}{\partial q}\left(\frac{\partial Y}{\partial y}\right) -
{\dfrac{1}{2\pi}}\frac{\partial F}{\partial y}\left(\frac{\partial Y}{\partial q}\right)\cong-
{\dfrac{1}{2\pi}}\frac{\partial F}{\partial y}\left(\frac{\partial Y}{\partial q}\right)\label{eq:ratio2}
\ee
that shows that the use of the approximation~(\ref{eq:sc.f.I_kind}) is justified in the $y$-range considered in our analysis.

\end{document}